\documentclass[fleqn,usenatbib]{mnras}

\usepackage{newtxtext,newtxmath}
\usepackage[T1]{fontenc}
\usepackage{ae,aecompl}

\usepackage{graphicx}	% Including figure files
\usepackage{amsmath}	% Advanced maths commands
\usepackage{amssymb}	% Extra maths symbols

\title[The DISPATCH Framework. I.]{DISPATCH: A Numerical Simulation Framework for the Exa-scale Era. I. Fundamentals}

\author[{\AA}. Nordlund, J. P. Ramsey, A. Popovas, M. K{\"u}ffmeier]{
{\AA}ke Nordlund,$^{1}$\thanks{E-mail: aake@nbi.ku.dk (\AA N)}
Jon P.\ Ramsey,$^{1}$\thanks{jramsey@nbi.ku.dk (JPR)}
Andrius Popovas$^{1}$
and Michael K{\"u}ffmeier,$^{1,2}$
\\
$^{1}$Centre for Star and Planet Formation, the Niels Bohr Institute and the Natural History Museum of Denmark,\\
University of Copenhagen, \O ster Voldgade 5-7, DK-1350 Copenhagen, Denmark\\
$^{2}$Zentrum für Astronomie, Heidelberg University, Albert Ueberle Str. 2, 69120 Heidelberg, Germany
}

\date{Accepted 2018 February 16. Received 2018 February 16; in original form 2017 June 26}

\pubyear{2017}

\begin{document}
\label{firstpage}
\pagerange{\pageref{firstpage}--\pageref{lastpage}}
\maketitle

\begin{abstract}
We introduce a high-performance simulation framework that permits the semi-independent, task-based solution of sets of partial differential equations, typically manifesting as updates to a collection of `patches' in space-time. A hybrid MPI/OpenMP execution model is adopted, where work tasks are controlled by a rank-local `dispatcher' which selects, from a set of tasks generally much larger than the number of physical cores (or hardware threads), tasks that are ready for updating. The definition of a task can vary, for example, with some solving the equations of ideal magnetohydrodynamics (MHD), others non-ideal MHD, radiative transfer, or particle motion, and yet others applying particle-in-cell (PIC) methods. Tasks do not have to be grid-based, while tasks that are, may use either Cartesian or orthogonal curvilinear meshes. Patches may be stationary or moving. Mesh refinement can be static or dynamic. A feature of decisive importance for the overall performance of the framework is that time steps are determined and applied locally; this allows potentially large reductions in the total number of updates required in cases when the signal speed varies greatly across the computational domain, and therefore a corresponding reduction in computing time. Another  feature is a load balancing algorithm that operates `locally' and aims to simultaneously minimise load and communication imbalance. The framework generally relies on already existing solvers, whose performance is augmented when run under the framework, due to more efficient cache usage, vectorisation, local time-stepping, plus near-linear and, in principle, unlimited OpenMP and MPI scaling.
% 
% 245 of 250 words
% 
\end{abstract}


%%%%%%%%%%%%%%%%%%%%%%%%%%%%%%%%%%%%%%%%%%%%%%%%%%

%%%%%%%%%%%%%%%%% BODY OF PAPER %%%%%%%%%%%%%%%%%%

\section{Introduction}
Numerical simulations of astrophysical phenomena are indispensable in furthering our understanding of the Universe. A particularly successful example is the many applications of the equations of magnetohydrodynamics (MHD) in the contexts of cosmology, galaxy evolution, star formation, stellar evolution, solar and stellar activity, and planet formation. As the available computer power has increased over time, so too has the fidelity and complexity of astrophysical fluid simulations; more precisely, one could say these simulations have consistently been at the limit of what is computationally possible. The algorithms and techniques used to exploit the available resources have also increased in ingenuity and complexity over the years; from block-based adaptive mesh refinement (AMR; \citealt{bergeroliger84,bc89}) to space-filling curves (e.g.\ Peano-Hilbert) to non-blocking distributed communication (e.g.\ Message Passing Interface version 3; MPI). The next evolution in high performance computing (HPC), `exa-scale'\footnote{Exa-scale computing is defined as the use of computer systems capable of $10^{18}$ floating point operations (FLOPs) per second.}, is approaching and, as currently available tools are quickly reaching their limits (see, e.g., \citealt{dubeyetal2014}), new paradigms and techniques must be developed to fully exploit the upcoming generation of supercomputers.

However, since exa-scale supercomputers do not yet exist, we instead choose to limit our definition of `exa-scale ready' to software that has intra- and inter-node scaling expected to continue to the number of cores required to reach exa-scales. Here, we present an intra-node task scheduling algorithm that has no practical limit, and we demonstrate that MPI-communications are only needed between a limited number of `nearby' nodes. The first property guarantees that we will always be able to utilize the full capacity inside nodes, limited mainly by the memory bandwidth. The second property means that, as long as cluster network capacity continues to grow in proportion to compute capacity, inter-node communication will never become a bottleneck. Load and communication balancing is only indirectly involved, in the sense that, for problems where static load balancing is sufficient, the scaling properties can be measured for arbitrary size problems. In what follows, we point out that balancing the compute load is essentially trivial when tasks can be freely traded between compute nodes, and that the most important aspect of load balancing then shifts to minimizing inter-node communications. We indicate how this can be accomplished using only communications with nearest neighbour nodes, leaving the details to a forthcoming paper.

There are many astrophysical fluid simulation codes currently available within the scientific community. The most commonly used techniques include single grids (e.g.\ 
ZEUS; \citealt{stonenorman1992_zeus,clarke1996}, 
FARGO3D; \citealt{benitezllambaymasset2016},
PLUTO; \citealt{mignoneetal2007_pluto}, 
ATHENA; \citealt{stoneetal08}, 
PENCIL; \citealt{brandenburgdobler2002_pencil}, 
Stagger; \citealt{nordlund+1994,kritsuketal2011}, 
BIFROST; \citealt{bifrost}), 
adaptively refined grids (e.g.\ 
ART; \citealt{kravtsovetal1997_art}, 
NIRVANA; \citealt{ziegler1998_nirvana},  
ORION; \citealt{klein1999_orion}, 
FLASH; \citealt{fryxelletal2000_flash}, 
RAMSES; \citealt{teyssier2002_ramses,fromangetal2006_ramses}, 
AstroBEAR; \citealt{cunninghametal2009_astrobear}, 
CASTRO; \citealt{almgrenetal2010_castro}, 
CRASH; \citealt{vanderholst2011_crash}, 
AZEuS; \citealt{rcm12}, 
ENZO; \citealt{bryanetal2014_enzo}, 
AMRVAC; \citealt{porthetal2014_amrvac}) 
and smoothed particle hydrodynamics (e.g.\ 
GADGET; \citealt{springel2005_gadget2}, 
PHANTOM; \citealt{lodatoprice2010_phantom}).
More recently, moving mesh and mesh-less methods have begun to emerge (e.g.\ 
AREPO; \citealt{springel2010_arepo}, 
TESS; \citealt{duffellmacfayden2011}, 
GIZMO; \citealt{hopkins2015_gizmo}, GANDALF; \citealt{hubberetal2018_gandalf}).
The range of physics available from one code to the next is very broad, and includes (but is not limited to) radiative transport, heating and cooling, conduction, non-ideal MHD, cosmological expansion, special and general relativity, multiple fluids, and self-gravity. 

There also exists a growing collection of frameworks which are solver- and physics-agnostic, but instead manage the parallelism, communication, scheduling, input/output and even resolution of a simulation (e.g.\ 
BoxLib\footnote{\url{https://ccse.lbl.gov/BoxLib/index.html}}, 
Charm++; \citealt{kaleetal2008_charm}, 
Chombo; \citealt{adamsetal2015_chombo}, 
Overture; \citealt{brownetal1997_overture}, 
Uintah; \citealt{berzinsetal2010_uintah}, 
PATCHWORK; \citealt{patchwork_2017}). 
Indeed, a few of these frameworks already couple to some of the aforementioned simulation codes (e.g.\ 
PLUTO+Chombo, 
CASTRO+BoxLib, 
ENZO-P/CELLO+Charm++).

Nearly all of the codes and frameworks mentioned above have fundamental weaknesses which limit their ability to scale indefinitely. In the survey of block-based adaptive mesh refinement codes and frameworks by \citet{dubeyetal2014}, the authors conclude that ``future architectures dictate the need to eliminate the bulk synchronous model that most codes currently employ''. Traditional grid-based codes must advance using time steps that obey the worst possible condition throughout the simulation volume, advancing at some fraction of a time step $\Delta t = \min (\Delta x/v_{\rm signal})$. The time step may be determined locally, but it is applied globally. For AMR codes with sub-cycling, where larger meshes can take longer time steps, these must generally be a multiple of the finer mesh time steps, resulting again in a global coupling of time steps. A global time step which is controlled by the globally worst single cell is problematic, and the problem unavoidably becomes exasperated as supercomputers grow larger and the physical complexity of models correspondingly increases. \citet{dubeyetal2014} also discuss challenges related to both static load balancing, where difficulties arise when multi-physics simulations employ physics modules with very different update costs, and dynamic load balancing, wherein the load per MPI process needs to be adjusted, for example, because of time-dependent adaptive mesh refinement.

Parallelism in both specific codes and frameworks is almost always implemented using (distributed-memory) MPI, shared-memory thread-based parallelism (e.g.\ OpenMP, pthreads), or a combination of the two. Although MPI has its limitations (e.g.\ fault tolerance support), and new approaches are beginning to emerge (e.g.\ partitioned global address space; PGAS), it is the de facto standard for distributed-computing and will likely remain so leading up to the exa-scale era. Therefore, efficient and clever use of MPI is necessary to ensure that performance and scaling do not degrade as we attempt to use the ever-growing resources enabled by ever-larger supercomputers (e.g.\ \citealt{wombat_2017}). An important aspect of this is efficient load balancing, whereby a simulation redistributes the work dynamically in an attempt to maintain an even workload across resources.

Herein, we put forward and explore novel techniques that not only enable better utilisation and scaling for the forthcoming exa-scale era, but immediately offers distinct advantages for existing tools. Key to this, among other features, are: The concept of locally determined time steps, which, in realistic situations, can lead to dramatic savings of computing time. Next, the closely related concept of task-based execution, wherein each task depends on only a finite number of `neighbouring' tasks. These tasks are often, but not necessarily, geometrically close. For example, sets of neighbouring tasks which solve a system of partial differential equations in space and time (which we denote as \emph{patches}) need to supply guard zone values of density, momentum, etc., to one another, but neighbouring \emph{tasks} can also encompass radiative transfer tasks with rays passing through the patch, or particle-based tasks, where particles travel through and interact with gas in a grid-based task. The neighbour concept can thus be described as a `dependency' concept. By relying on this concept, we can ensure that MPI processes only need to communicate with a finite number of other MPI processes and, thus, we can, to a large (and often complete) extent, avoid the use of MPI global communications. In this regard, the neighbour concept provides the potential for essentially unlimited MPI scalability. The avoidance of global operations and the neighbour concept also applies to intra-node shared-memory parallelism: First, we keep the number of OpenMP `critical regions' to an absolute minimum and instead rely on `atomic' constructs. Second, we employ many more tasks per MPI rank than there are hardware threads, ensuring there are many more executable tasks than neighbour dependencies. Intra-node scalability is therefore, in general, limited only by memory bandwidth and cache usage.

These features are the foundation of the DISPATCH simulation framework. In what follows, we first describe the overall structure of the framework, outlining its constituent components and their interaction (Section \ref{sec:hierarchy}). In Section \ref{sec:perspectives}, we then review the code structure in the context of execution of an experiment. In Section \ref{sec:components}, we describe the major code components of the DISPATCH framework in detail. In Section \ref{sec:examples}, we validate the concepts and components by using four different numerical experiments to demonstrate the advantages of the DISPATCH framework and confirm that the solvers employed produce the same results as when used separately. To emphasise the framework aspect, each of the experiments uses a different solver; two internal solvers (a hydrodynamical Riemann solver and a staggered-mesh RMHD solver) and one external solver (ZEUS-3D, implemented as an external library). Finally, in Section \ref{sec:summary}, we summarise the properties and advantages of the DISPATCH framework, relative to using conventional codes such as RAMSES, Stagger, and AZEuS. In follow-up work (Popovas et al.; Ramsey et al., in prep.), we will extend our description of the DISPATCH framework to include dynamic refinement, moving patches and patches of mixed coordinate types.

\section{Object Hierarchy}
\label{sec:hierarchy}
DISPATCH is written in object-oriented Fortran and relies heavily on the concept of inheritance\footnote{In Fortran, `objects' are called `derived types', and inheritance is implemented by `extending' a derived type.}. Certainly, the ideas and concepts within DISPATCH could easily be carried over to another object-oriented language, such as C++, but conversely, Fortran does not impose any serious language-related limitations on their implementation. 

The framework is built on two main classes of objects: \textit{tasks} and \textit{task lists}.

\subsection{Tasks}
\label{sub:tasks}
The task class hierarchy has, as its defining member, a
\begin{enumerate}
\item \textit{task} data type, which carries fundamental state information such as task position, times, time steps, the number of time slices stored, status flags, rank, and so on. The \textit{task} data type also includes methods\footnote{`Type-bound procedures' in Fortran speak.} for acquiring a task ID, setting and inquiring about status flags, plus `deferred methods' that extending objects must implement. All mesh-based tasks extend the \textit{task} data type to a
\item \textit{patch} data type, which adds spatial properties, such as size, resolution, number of guard zones, coordinate system (Cartesian or curvilinear), number of physical variables, etc. Patches also contain methods for measuring intersections between different patches in space and time, as well as for writing and reading snapshots. These are generic properties and methods, and are shared by all mesh-based 
\item \textit{solver} data types, which specify the physical variables to be advanced, adds methods to initialise and advance the patch data forward in time, plus any parameters that are specific to the solver in question. The specific solver data type is then extended to an
\item \textit{experiment} data type, which adds experiment-specific functionality, such as initial and boundary conditions, in addition to the specific update procedure for the experiment. The \textit{experiment} data type also serves as a generic wrapper that is accessed from the \textit{task list} class hierarchy, effectively hiding the choice of solver, thus making it possible to execute the same experiment with different solvers.
\end{enumerate}

\subsection{Task lists}
\label{sub:tasklists}
The task list class hierarchy has as its base member a 
\begin{enumerate}
\item \textit{node} data type, which defines a single node in a doubly-linked list that points to, and carries information about, individual tasks. In addition to pointers to the previous and next nodes, nodes have a pointer to the head of a linked list of \textit{nbors}, a concept that generalises neighbours beyond spatial proximity to include tasks that in one way or another depend on the current task, or because the current task depends on them. The \textit{node} data type also includes methods to initialise and maintain its own neighbour list. Individual nodes are used by the
\item \textit{list} data type, which defines a doubly-linked list of nodes and keeps track of its properties. The \textit{list} data type also contains the generic methods necessary to manipulate linked lists, such as appending, removing and sorting of nodes. It is, however, the
\item \textit{task list} data type that extends the \textit{list} data type with methods that are specific to the execution of tasks and the handling of task relations. In particular, the \textit{task list} data type contains the \textit{update} method, which is a key procedure in DISPATCH and responsible for selecting a task for updating. The complete execution of a DISPATCH experiment essentially consists of calling the task list \textit{execute} procedure, which calls the task list \textit{update} procedure repeatedly until all tasks are finished; this is typically defined as having advanced to a task's final time.
\end{enumerate}

\subsection{Components}
\label{sub:components}
The primary means of generating task lists in DISPATCH is via a set of \textit{components}, each of which produces a subset of tasks, organised in some systematic fashion. For example, one of the most frequently-used components is the \textit{cartesian} component, which generates and organises tasks in a non-overlapping, regular and Cartesian-like spatial decomposition; in this case, each task initially has, in general, $3^3-1 = 26$ spatial \textit{nbors}. The corresponding mesh-based tasks/patches may, in turn, use Cartesian or orthogonal curvilinear coordinates; in the latter case, the Cartesian-like partitioning of tasks is performed in curvilinear space.

Another commonly-used component creates nested sets of `Rubik's Cube' (3x3x3) or `Rubik's Revenge' (4x4x4) patches. For example, one can arrange 27 patches into a 3x3x3 cube, and then repeat the arrangement recursively by splitting the central patch into 3x3x3 child patches, each with a cell size that is three times smaller than its parent patch. A `Rubik's Revenge' setup is analogous, except that it uses 4x4x4 patches and the central 2x2x2 patches are each split into a new 2x2x2 arrangement, leading to a resolution hierarchy with a factor of two decrease in patch and cell size with each additional level. The nested set can furthermore be complemented by repeating the coarsest level configuration in one or more directions.

The Rubik's-type component is useful, for example, to represent the environment near a planetary embryo embedded in an accretion disk, making it possible to cover the dynamic range from a small fraction of the planet radius to the scale height of the accretion disk with a relatively limited number of patches (e.g. [no.\ levels + no.\ extra sets]$\times 27$). The set of patches produced by a Rubik's component can then be placed in the reference frame co-moving with the planetary embryo, with corresponding Coriolis and net inertial (centrifugal) forces added to the solver, thus gaining a time step advantage relative to the stationary lab frame.

Another kind of component available in DISPATCH, exemplified by sets of tasks that solve radiative transfer (Sect.\ \ref{sub:rad_trans}), may overlap spatially with patches, but there is additionally an important \textit{causal} dependency present. For example, patches which solve the equations of MHD provide densities and temperatures to sets of radiative transfer tasks in order to calculate the radiation field. In return, the radiative transfer tasks provide heating or cooling rates to be applied to the MHD patches prior to the next update. This kind of `causally-linked' component can operate either as an extension on top of another task, or as a semi-independent set of tasks coupled only by pointers. In the former, the task presents as only a single task and is therefore updated in step with its causal \textit{nbor}. In the latter, there is substantial freedom in choosing when to update the component. A good example of this in practice in BIFROST \citep{bifrost} solar simulations, where the radiative transfer problem is solved with a slower cadence than the one used to evolve the MHD \citep{bifrostRT}.

\subsection{Scenes}
\label{sub:scenes}
Components, such as those described in the previous subsection, are used as building blocks to create a \textit{scene} hierarchy, where one could have, for example, a top level that is a galaxy model, which, in its spiral arms, contain a number of instances of giant molecular cloud components, each of which contains any number of protostellar system components, where a protostellar system component contains an accretion disk component, with a collection of moving patches representing the gas in the accretion disk. These patches would be orbiting a central star, whose evolution could be followed by, for example, a MESA (\citealt{paxton2010}) model, which takes the accretion rate directly from a sink particle component that represents the star. In addition, the protostellar system component could contain Rubik's Cube components, co-moving with planetary embryos, which could, in turn, be coupled to particle transport tasks representing dust in the accretion disk. Indeed, scenes are the way to build up simulations, complex or otherwise, from one or more individual components in DISPATCH; in the end, a scene need only provide a task list that is ready for execution, i.e., with the nbor relations already determined.

\section{Code Functionality}
\label{sec:perspectives}
The overall structure and functionality of the code framework is best understood by combining several possible perspectives on the activities that take place as an experiment is executed. Below we adopt, one-by-one, perspectives that take A) the point of view of a single task, and the steps it goes through cyclically, B) the point of view of the task scheduler/dispatcher, as it first selects tasks for execution/updating and then later checks on these tasks to evaluate the consequences of the updates. Then C) we take the point of view of the ensemble of MPI processes, and their means of communicating. Next, D) we take the point of view of the load balancer, and look at how, in addition to keeping the load balanced between the MPI processes, it tries to minimise the actual need to communicate. Finally, E) we take the point of view of the input/output (I/O) sub-system, and examine how, within the ensemble of MPI processes, snapshots can be written to disk for post-processing and/or experiment continuation.

\subsection{A) Single task view}
\label{sub:singletask}
For simplicity, we consider the phases that a mesh-based \textit{experiment} task goes through; the sequence that a different type of task experiences does not differ substantially. An \textit{experiment} task, an extension of a \textit{solver} task, itself an extension of a \textit{patch} task, relies on having guard zone values that are up-to-date before it can be advanced to the next time step. In general, guard zone values have to be interpolated in time and space, since patches use local time-stepping (and are therefore generally not synchronised in time) and can have differing resolutions. To enable interpolation in time, values of the field variables (e.g.\ density, momentum, etc.) in each patch are saved in a number of time slices (typically 5--7) using a circular buffer. Interpolation in space, prolongation, and its inverse process, restriction, are meanwhile accomplished using conservative interpolation and averaging operators.

The `dispatcher' (see below) checks if neighbouring patches have advanced sufficiently enough in time to supply guard zone values to the current patch before moving it to a (time-sorted) `ready queue'. After having been selected by the dispatcher for update, patches are then switched to a `busy' state, in which both the internal state variables (e.g.\ the MHD variables) and the patch time is updated. The new state overwrites the oldest state in the circular buffer, and since guard zone values corresponding to the next time step are not yet available, the task is put back in to the `not ready' state. Occasionally, when the patch time exceeds its next scheduled output time, an output method (which can be generic to the \textit{patch} data type or overloaded by a \textit{solver} or \textit{experiment}-specific method) is called (see below).

\subsection{B) Task scheduler view}
\label{sub:taskscheduler}
In the series of states and events discussed above, the selection and preparation of the task for update is the responsibility of the task scheduler \textit{update} method, which is a central functionality (in practice, spread over several methods) in the DISPATCH code -- this is essentially the \textit{dispatcher} functionality that has given the code framework one of the inspirations for its name (the other one derived from its use of partially DISconnected PATCHes).

The task list update procedure can operate in two modes. It can either
\begin{enumerate}
\item let threads pick the oldest task from a linked list of tasks (the ready queue) that have been cleared for update. This linked list is sorted by time (oldest first), so a thread only has to pick off the head from the queue and execute its update method. In this scheme, after updating the task, the thread immediately checks the neighbours of the updated task to see if perhaps one of those tasks became `ready', e.g.\ because the newly updated task is able to provide the last piece of missing guard zone data. If any `ready' tasks are found, they are inserted into the `ready queue' in ascending time order. Alternatively,
\item the task list update procedure runs on a single OpenMP thread, picks off the oldest task from the ready queue and subsequently spawns a thread to update it. As before, it then searches the neighbour list for tasks that possibly became executable, and adds these tasks to the ready queue.
\end{enumerate}
Note that, regardless of the operational mode, the task scheduler is rank-local.

The first operational mode is the simpler one; indeed, thread-parallelism in this mode is implemented using a single \texttt{\$omp\,\,parallel} construct placed around the task list update method. This mode, however, has the drawback that two of the linked list operations -- picking off the head and adding new tasks to the ready queue -- are continuously being performed by a large number of threads in parallel. These two operations must therefore be protected with OpenMP critical regions to ensure that only one thread at a time is allowed to manipulate the ready queue. This is not a problem as long as the number of threads per MPI process is limited; even on a 68-core Intel Xeon Phi processor, there is hardly any measurable impact.

However, in order to obtain truly unlimited OpenMP scalability, it is necessary to operate entirely without OpenMP critical regions, which is possible with the second mode of operation. In this mode, a single (master) thread is responsible for both removing tasks from the ready queue, as well as adding new tasks to it. In this mode, the actual task updates are performed by OpenMP threads initiated by \texttt{\$omp task} constructs, with the task list update method running on the master thread being responsible (only) for spawning tasks and handling linked list operations. Therefore, no critical regions are needed, and one can reach much larger numbers of threads per MPI process. The load on the master thread is expected to be ignorable, and if it ever tended to become noticeable, it can easily spawn sub-tasks that would take care of most of the actual work. Note also, that the default behaviour of OpenMP task constructs is that the spawning task can also participate in the actual work (possibly encouraged to do so by suitably placed \texttt{\$omp\,\,taskyield} constructs).

\subsubsection{Defining task `readiness'}
\def\isahead{{\tt is\_ahead\_of}\ }
The definition of when a task is considered ready for updating is intentionally flexible. By default, it is implemented via a logical function \isahead which examines the difference between the time of a task and its neighbours. A particular task (``{\tt self}'') is deemed ready to be updated if the condition,
\begin{equation}
t_{\rm self} \le t_{\rm nbor} + g \Delta t_{\rm nbor},
\end{equation}
is true for all tasks in the neighbour list of {\tt self}. Here, $g$ is a `grace' parameter which specifies the amount of extrapolation permitted relative to the neighbour time step, $\Delta t_{\rm nbor}$. Note that, since the actual update will take place later and other tasks are constantly being updated by other threads, even though an extrapolation is allowed, it is not necessarily required when the actual update happens. From experience, setting $g=0.05$ already generally increases the number of tasks in the ready queue at any given time significantly, while, on the other hand, setting $g$ as a high as 0.3 has not been found to produce visible glitches in results.

The \isahead function may be overloaded at any level of the task hierarchy. This may be used, e.g., where the tasks of a particular experiment are defined. One might decide, as is done routinely in the BIFROST code, that the radiative transfer solution should be calculated less frequently than the dynamics, and thus \isahead should be overloaded to return true when the causal neighbours of {\tt self} satisfy an appropriate criterion.

When extrapolation in time is actually used, as well as when interpolating in time to fill guard zone values, the default action is to use linear inter- and extrapolation in time. However, since several time slices are available, one may choose to use higher order time inter- and extrapolation. Second-order extrapolation in time has, e.g., turned out to be optimal for the gravitational potential when solving the Poisson equation under certain circumstances \citep{ramsey+2018}.

\subsection{C) MPI process view}
\label{sub:mpiview}
In DISPATCH, there is typically one MPI process per compute node or per physical CPU (central processing unit) `socket'. Each MPI process also typically engages a number of OpenMP threads that matches or exceeds (if `hyper-threading' is supported and favourable) the number of physical cores. Since the number of tasks that are executable at any given time may be a small fraction of the total number of tasks, each MPI process typically `owns' a number of tasks that significantly exceeds the number of available OpenMP threads. Some of these tasks are naturally located near the geometrical edge of a rank (henceforth `boundary tasks'), and they therefore could have neighbour tasks that belong to a different MPI process (henceforth `virtual tasks'); the remaining `internal' tasks have, by definition, neighbours that are entirely owned by the same MPI process.

As soon as a task marked as a boundary task has been updated by an OpenMP thread, the thread prepares an MPI package and sends it to all MPI processes that need information from that task (i.e.\ its `nbors'). Conversely, each time a thread receives a package from a nbor, it immediately issues a new \texttt{MPI\_IRECV} to initiate receipt of the next package. Threads that become available for new work start by checking a share of the outstanding message requests using a thread private list, which eliminates the need for OpenMP critical regions when receiving and unpacking MPI packages.

The MPI packages are typically used for supplying guard zone values. By sending this information pro-actively (using \texttt{MPI\_ISEND}), rather than having other MPI processes ask for it, the latency can be greatly reduced, and chances improve that the boundary patches of other processes will have up-to-date data available in their virtual neighbours to supply guard zone data when their neighbour lists are checked.

The MPI packaging for patches contains both a copy of the most relevant task parameters and the state variables in the interior of the patch. There are essentially two reasons to ship data for all interior zones, rather than limiting the package content only to layers that will be needed for guard zone data. First, this simplifies package creation and handling, and network capacity is generally large enough to allow this for moderately-sized patches without any noticeable slowdown. Second, this simplifies load balancing (see below), since it allows changing the owner of a patch by simply `giving' a patch to a neighbour MPI process. The patch then needs only change its status, from `virtual' to `boundary' on the receiver side, and from `boundary' to `virtual' on the sender side.

Should memory bandwidth actually become a problem one can reduce the network traffic significantly by only sending guard zone values, at the cost of added complexity in the package pack and unpack methods and in the load balancing methods.

\subsection{D) Load balancer view}
\label{sub:loadbalancer}
Given the ownership swap protocol outlined above, the actual balancing of workload is a nearly trivial task, since any MPI process that finds itself with a surplus of work, relative to a neighbouring MPI process, can easily `sell' some of its boundary tasks to its less-loaded neighbour MPI processes. However, doing so indiscriminately risks creating ragged geometric boundaries between neighbouring MPI processes. In this sense, the load balancer should also function as a `communication balancer' which attempts to keep the number of MPI communications per rank and per unit simulation time near some minimum value.

Load balancing follows each task update, and can be performed after every update, after a given number of updates of each task, or after a given wall clock time has passed since the previous load balance. Load balancing is entirely local, with each task evaluating (in an OpenMP critical region) if the number of communications, or the load imbalance, can be reduced by passing ownership of the task to one of its rank neighbours. A minimum threshold for imbalance is implemented to prevent frequent passing of tasks that only marginally improves the imbalance. Only boundary tasks, i.e., tasks that have neighbours on other ranks are candidates for ownership transfer. If the transfer is beneficial, the task indicates to the rank neighbour that ownership should be changed. The rank that takes over ownership then changes the task status from `virtual' to `boundary', while the previous owner changes the status from `boundary' to `virtual'. All affected tasks then refresh their neighbour lists to reflect the new situation.

By using emulation of an initially and severely fragmented patch distribution via software prototyping, we find that it is more efficient to address the load imbalance and the communication reduction in two separate steps while, within each step, the other aspect of the balancing is more or less ignored.

The procedure is essentially as follows:
\begin{enumerate}
\item Count the number of MPI neighbours for each rank by enumerating the number of different ranks recorded in the neighbour lists of a rank's boundary tasks;
\item Evaluate, for each boundary task, if the total number of MPI neighbours on the involved ranks would be reduced if ownership of a task is swapped. In this step, one carries out the most advantageous swaps, even if a swap would tend to introduce a load imbalance.
\item In a separate step, evaluate if any task swaps can be made that restore load balance, without increasing the number of MPI neighbours.
\end{enumerate}

The reason this two step procedure is advantageous, relative to trying to reduce communications under the simultaneous constraint to avoid causing imbalance, is that it is relatively difficult to find good swaps in step (ii), and the process is made easier if load (im)balance is temporarily ignored. It is, meanwhile, relatively easy in step (iii) to find swaps that improve load balance without changing the total number of of MPI neighbours.

\subsection{E) Input/Output view}
\label{sub:ioview}
After a task has been updated, its current time in code units is compared to the \texttt{task\%out\_next} parameter and, upon passage of that \texttt{out\_next}, the \textit{task} output method is called and a snapshot is written. The actual method invoked depends on where in the task hierarchy the generic output method has been overridden. For example, if an \textit{experiment}-specific method has not been implemented, then a \textit{solver}-specific method will be invoked if implemented, otherwise a generic \textit{patch} output method is used. Currently, DISPATCH snapshots use one of two output formats: One that writes patch data as raw binary data to one file, and patch information as text to another file. An alternative output format uses the packing procedure used to  communicate task data between MPI ranks, and collects the data for all tasks on a MPI rank into a single file. In either case, these snapshots are suitable not only for visualisation but also for restarting a simulation. Finally, for reasons of portability and performance, parallel HDF5 support is currently being explored.

\section{Current code components}
\label{sec:components}
\subsection{Internal and external HD, MHD, and PIC solvers}
\label{sub:pdesolvers}
The solvers from a few well-used and well-documented astrophysical fluid codes have been ported to the DISPATCH framework and validated using experiments carried out with both the original code and the DISPATCH implementation. As a first representative of Godunov-type Riemann solvers, the HLLC solver from the public domain RAMSES code \citep{teyssier2002_ramses} has been incorporated. To take advantage of the DISPATCH speed advantages in stellar atmosphere and similar work (e.g.\ \citealt{baumannetal2013}), several versions of the Stagger Code \citep{nordlund+1994,kritsuketal2011} class of solvers have also been ported. As an example of connecting to an external solver, the ZEUS-3D solver \citep{clarke1996,clarke2010} used in the AZEuS adaptive mesh refinement code \citep{rcm12} is available; the ZEUS-3D solver is also the only currently-included solver that can leverage the advantage of orthogonal, curvilinear coordinate systems (e.g.\ cylindrical, spherical).

Incorporating an external solver as a library call in DISPATCH is straightforward, but requires some modification of the external code. An explicit interface must be defined by the external solver. One must also ensure that the external solver is `thread-safe', i.e., it can safely be invoked by many threads at once. Initialisation and updating procedures are meanwhile defined in the DISPATCH {\it solver} module; these procedures exist for all solvers in DISPATCH, external or otherwise. As part of the update procedure, conversion procedures may be required to ensure the physical variables are in a format suitable for the solver. For example, in ZEUS-3D, velocity is a primary variable while, in DISPATCH, we store the momentum. After the external solver is called, the data must be converted back to DISPATCH variables. At this point, the developer decides whether DISPATCH or the external solver determines the next time step. In the case of ZEUS-3D, it determines the next Courant-limited time step internally and returns the value to DISPATCH. Calling an external solver occurs at the same point in DISPATCH as any other solver, as part of a \textit{task} update procedure. Memory management within the external solver is, meanwhile, not controlled by DISPATCH.

In ongoing work, we are also incorporating the public domain version of the photon-plasma particle-in-cell code, PPcode\footnote{\url{https://bitbucket.org/thaugboelle/ppcode}} \citep{haugbolleetal2013_ppcode}, and the BIFROST MHD code \citep{bifrost}, including its modules related to chromospheric and coronal physics. We intend to use these solvers for multiple-domain-multiple-physics experiments in the context of modelling solar and stellar atmospheric dynamics driven by sub-surface magneto-convection, expanding on the type of work exemplified in \citet{baumannetal2013}.

\subsection{Radiative transfer}
\label{sub:rad_trans}
Radiative processes are undeniably important in most astrophysical applications (e.g.\ planetary atmospheres, protoplanetary disk structure, HII regions, and solar physics, to name a few). Being a 7-dimensional problem (three space + time + wavelength + unit position vector), it can be a daunting physical process to solve. A number of approaches have been developed over the last several decades in an attempt to tackle the problem, including flux-limited diffusion (FLD; \citealt{levermorepomraning1981_fld}, Fourier transforms (e.g.\ \citealt{cen2002}), Monte Carlo techniques (e.g.\ \citealt{robitaille2011}), variable tensor Eddington factors \citep{dullemondturolla2000_vtef}, short (e.g.\ \citealt{stonetal1992}), long (e.g.\ \citealt{nordlund82,Heinemann2006}) and hybrid \citep{rijhorstetal2006_hybrid} characteristics ray-tracing. 

DISPATCH currently implements a hybrid-characteristics ray-tracing radiative transfer (RT) scheme: long rays inside patches and short rays in-between, albeit with a few exceptions as described below. The RT module consists of a ray geometry component, an initialisation component, and run time schemes; each component needs to know very little about how the others work, even though they rely heavily on each other. In addition, since the RT module depends on the values of, for example, density and temperature, it also relies on the co-existence of instances of a \textit{solver} data type that can provide these quantities. Below, we briefly explain the purpose of each of the RT components; a detailed description of the RT module, together with implementation details for refined meshes and additional validation tests, will appear in a subsequent paper of this series (Popovas et al., in prep.). 

Current solver implementations include Feautrier and integral method formal solvers (e.g.\ \citealt{nordlund1984}), multi-frequency bin opacities, and source function formulations with either pure thermal emission, or with a scattering component\footnote{While the 2nd order Feautrier method conserves energy by construction, the more accurate integral method does not. As such, the choice of RT solver should be made with this in mind.}. Scattering is handled in a similar way as in the BIFROST code \citep{bifrostRT}. For cases with low to moderate albedos, the method is essentially lambda-iteration. The specific intensities from previous time steps are stored in DISPATCH time slices and are used as starting values for the next time steps, which makes this approach effective. Extreme scattering cases could be handled by implementing more sophisticated methods, such as accelerated lambda iteration \citep{Hubeny2003}.

\subsubsection{Ray geometry}
When the RT module is initialised, it first creates a ray-geometry (RG) for existing patches. The only information the RG component needs is the simulation geometry type (Cartesian/cylindrical/spherical), dimensions of a patch (number of cells per direction), the desired number of ray-directions, and their angular separation. The RG component then proceeds to spawn rays: Each ray starts at one patch boundary and ends at another boundary (face, edge, or corner) of the same patch. Spacing of rays is set to the patch cell size along the direction that forms the largest angle with the ray. The ray points are, ideally, co-centred with patch cell centres along the direction of the ray. In this case, the required hydrodynamic quantities (e.g.\ density and internal energy) do not need to be interpolated to the RG, and neither do the resulting intensities need to be interpolated back to the mesh; this saves a significant amount of computational time.  If angular resolution surpassing that provided by rays along axes, face diagonals and space diagonals is required, this can be accomplished via rays in arbitrary directions but with an extra cost resulting from interpolating temperature and density to ray points and then interpolating the resulting net heating or cooling back to the patch mesh. In this case, as with the MHD variables, interpolation to ray points and of the subsequent heating rate back to the mesh are done using conservative operators.

\begin{figure}
  \includegraphics[width=1.0\columnwidth]{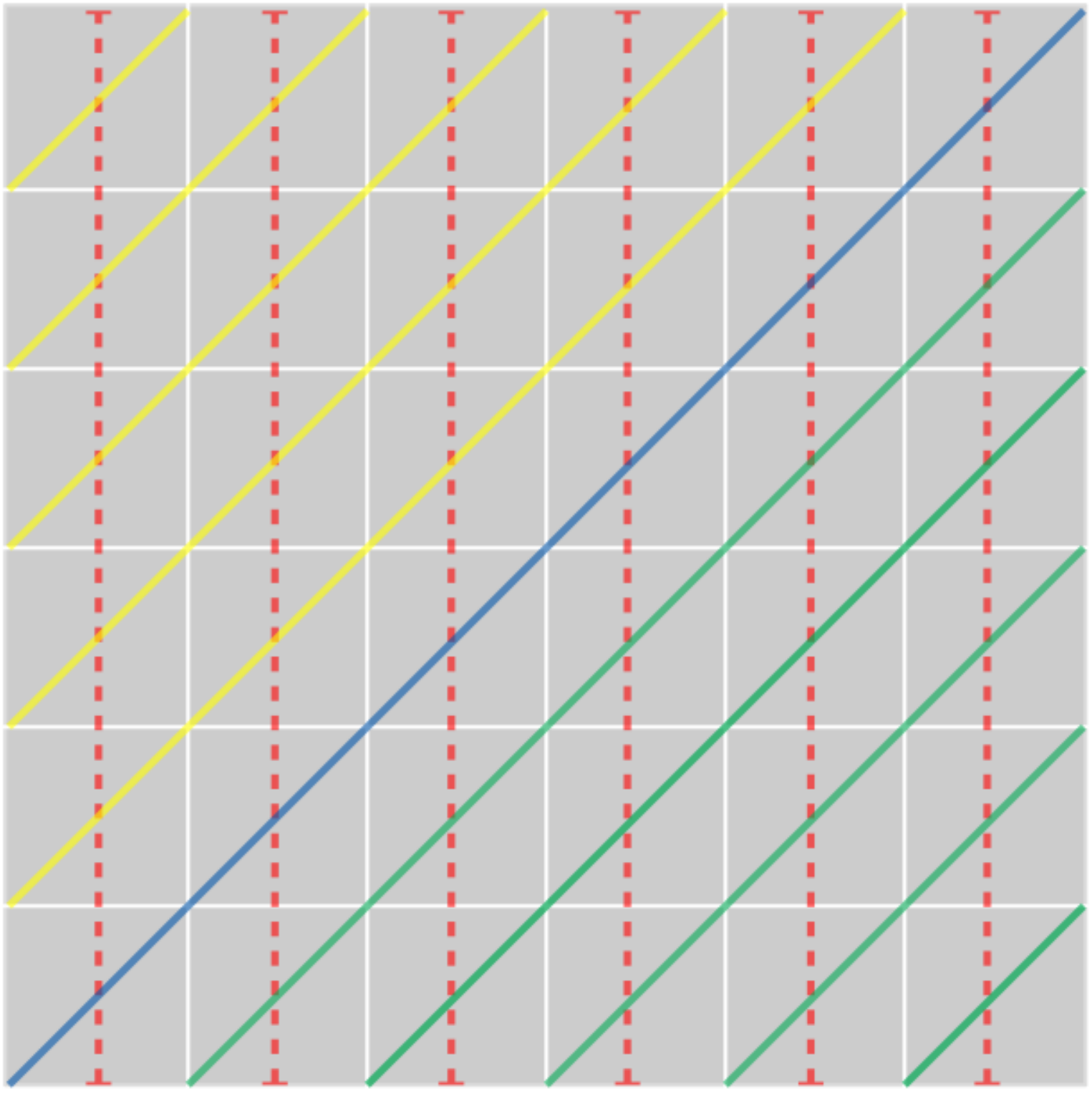}
  \caption{Ray geometry representation in 2-D. (M)HD grid cells are represented by grey squares, coloured lines represent rays. Different line styles represent different directions. Different colours represent different bundles.}
  \label{fig:ray_geom}
\end{figure}

Once all rays have been created, they are organised into a fast look-up tree hierarchy. The primary access to this tree is through the ray geometry type. Within a \textit{ray geometry} object, the rays are further sub-divided into
\begin{enumerate}
\item \textit{ray directions} - in principle, as long as scattering is not considered, rays in one direction do not care about rays with another direction, so they can be separated and updated as independent tasks by different threads. To solve for RT inside a patch, boundary conditions from the patch walls (i.e.\ boundary face) that the rays originate from are required. Not all slanted rays with one direction originate and terminate at the same wall (see Figure \ref{fig:ray_geom}). To avoid waiting time originating from sets of \textit{ray directions} that end on more than one patch wall, the complete set is further sub-divided into
\item \textit{ray bundles}; these are defined as sets of rays that originate and terminate on the same pair of patch walls. RT is a highly repetitive task, where a large numbers of rays is typically considered. It is thus highly advantageous to use schemes that maximise hardware vectorisation in order to reduce the computation time. Therefore, all rays should, preferably, be of the same length. This is a natural feature of rays that are parallel to a patch coordinate axis; slanted ones, meanwhile, are further rearranged into
\item \textit{ray packets}; these are the sets of rays in one \textit{ray bundle} that have the same length. To promote faster data lookup (e.g.\ interpolating ray coordinates to patch coordinates, origin/termination points, etc.), they are organised as data arrays. 
\end{enumerate}

Note that axis-oriented rays, as well as rays with specific spatial angles (e.g.\ 45$^{\circ}$), will generally terminate at locations that coincide with the origins of rays in a `downstream' (with respect to RT) patch. This is exploited in the current implementation by introducing the concept of \textit{ray bundle chains}.

\subsubsection{Ray bundle chains}
To obtain the correct heating/cooling rates inside a patch, one needs boundary values/incoming intensities from `upstream' patches. By reorganising rays into bundles (see above), one minimises the number of walls needed per particular bundle down to a single pair: one at the origin and one at the termination. Such an arrangement allows the connection of one ray bundle to a similar ray bundle in a neighbour patch, resulting in a \textit{bundle chain}. If we continue the analogy with actual chains, then the first, or master, chain `link' is connected to an outer boundary, and all the other links, by virtue of the chain, obtain their boundary values from their upstream link immediately after the RT is solved on the upstream link. If the upstream surface of a link is a physical boundary or a change in spatial resolution, this particular link is defined as being the first link in the chain, i.e., the one from which execution of the chain starts. Bundle chains effectively transform the short-characteristic ray scheme inside patches into long-characteristic rays over many patches along the ray bundle direction.

\subsubsection{RT initialisation and execution}
After the ray geometry has been generated, the RT initialisation component steps through all of the patches in the task list and creates an overlapping RT patch. The RT patch keeps track of all the ray-directions and sets up bundle chains by searching for upstream patches in a patch's neighbour list. 

When an (M)HD patch, and subsequently the internal state variables, is updated, a corresponding RT patch is ready to be executed. Initially, it may not have wall/boundary data available, but it can still prepare the internal part by evaluating the source function and opacities for the internal cells. The RT patch then steps through its list of bundles and `nudges' the master links of any bundle chains. The master link could belong to this patch, or it could belong to a very distant patch along a particular ray direction. In most cases, chain links along an arbitrary ray direction are not yet ready for execution (e.g.\ the internal data for an RT patch is not prepared yet), so the master link simply records which links in the chain are ready. As soon as a last nudge indicates that all links are ready, the chain can be executed recursively, as follows:
\begin{enumerate}
\item The first bundle in the chain gets boundary values either from physical boundary conditions, or retrieves them from an upstream patch; in the latter case, the upstream source is either a virtual patch on another MPI rank or has a different spatial resolution (and therefore requires interpolation);
\item Simultaneously, the source function and opacities are mapped from internal patch data onto the ray bundle coordinates;
\item The boundary values and mapped internal data are sent to an RT solver. The solver returns the heating/cooling rates for internal cells in addition to wall intensities, which, in turn, are sent to the next, downstream link in the bundle chain.
\end{enumerate}

Once a bundle chain has executed, the chain and its individual links are placed in a `not ready' state while the chain waits for the corresponding MHD patches to be updated again. In practice, it may not be necessary (or practical) to solve RT after every MHD update. Thus, one can instead specify the number of MHD updates or time interval after which the next RT update will be scheduled.

\subsection{Non-ideal MHD}
\label{sub:non-ideal_MHD}
The importance of magnetic fields in astrophysics is well-established, as is the application of MHD to problems involving magnetised fluids. However, non-ideal MHD effects (Ohmic dissipation, ambipolar diffusion, and the Hall effect) can be important in various astrophysical contexts, for instance, in protoplanetary disks and the interstellar medium.  Here, we account for the additional physical effects of Ohmic dissipation and ambipolar diffusion in DISPATCH by extending the already implemented MHD solvers. We illustrate the process using the staggered-mesh solvers based on the Stagger Code \citep{nordlund+1994,kritsuketal2011}; the implementation in other solvers would differs mainly in details related to the centring of variables. In the Stagger and ZEUS-3D solvers, for example, components of momentum are face-centred, while in the RAMSES MHD solver, momentum is cell-centred. In all of the currently implemented MHD solvers, electric currents and electric fields are edge-centred, following the constrained transport method \citep{evanshawley1988} to conserve $\nabla\cdot\bf B = 0$.

\subsubsection{Ohmic dissipation}
Ohmic dissipation results from the additional electromotive force (EMF) induced by imperfect conduction (non-zero resistivity) in a magnetised fluid. For simplicity, we assume an isotropic resistivity that can be described by a scalar parameter, $\eta_{\rm Ohm}$, and write the additional electric field as:
\begin{equation}
	\label{eq:Ohm}
    \mathbf{E}_{\rm Ohm}  = \eta_{\rm Ohm} \mathbf{J},
\end{equation}
where $\mathbf{J} = \nabla\times\mathbf{B}$ is the current density and $\mathbf{B}$ is the magnetic field; $\mathbf{E}_{\rm Ohm}$ is added to the ideal EMF ($\mathbf{E} = -\mathbf{v}\times\mathbf{B}$) prior to invoking the induction equation. Meanwhile, a non-zero resistivity also results in an additional heating term:
\begin{equation}
	\label{eq:Ohm_heat}
	Q _{\rm Ohm} = \mathbf{J}\cdot \mathbf{E}_{\rm Ohm}  = \eta_{\rm Ohm} J^2 .
\end{equation}
Since the current density is edge-centred, we must appropriately average its components to compute the cell-centred heating rate.

Additionally, we use the following expression for the Ohmic time step:
\begin{equation}
\label{eq:dtOhm}
\Delta t_{\rm Ohm} = C _{\rm Ohm}\cdot \min \left(\frac{\Delta x^2}{\eta_{\rm Ohm}} \right),
\end{equation}
where $\Delta x$ is the cell size and $C_{\rm Ohm}$ is a Courant-like parameter, conservatively chosen to be $0.1$ following \citet{Massonetal2012}. The Courant time step is subsequently modified to select the minimum of $\Delta t_{\rm Ohm}$ and the usual $\Delta t_{\rm MHD}$.

\subsubsection{Ambipolar diffusion}
In the case of a gas that is only partially ionised, magnetic fields directly affect the ionised gas but not the neutral gas; the neutral gas is instead coupled to the ions via an ion-neutral drag term. In general, therefore, the ionised gas moves with a different speed than the neutral gas. If the speed difference is small enough to neglect inertial effects, the drag term can be represented by estimating the speed difference (i.e.\ the `drift speed') between the ionised and neutral gas. We implement this so-called single-fluid approach along the lines of previous works (cf.\ \citealt{maclow+1995, Padoanetal2000, duffinetal2008, Massonetal2012}). 

Specifically, we add an additional ambipolar diffusion EMF, 
\begin{equation}
\label{eq:AD}
\mathbf{E}_{\rm AD} = -\mathbf{v}_D \times \mathbf{B},
\end{equation}
to the induction equation, where
\begin{equation}
\mathbf{v}_D = \frac{1}{\gamma_{\rm AD}\rho \rho_i} \mathbf{J} \times \mathbf{B}
\end{equation}
is the drift velocity, $\rho$ is the total mass density, $\rho_i$ is the ion mass density and $\gamma_{\rm AD}$ is the ambipolar drift coefficient. 

Like Ohmic dissipation, ambipolar diffusion also results in an additional heating term:
\begin{equation}
Q_{\rm AD} = \rho \rho_{i} \gamma_{\rm AD} (\mathbf{v}_D)^2.
\end{equation}
As before, we appropriately average the components of $\mathbf{J}\times\mathbf{B}$ to cell-centre in order to compute the cell-centred $Q_{\rm AD}$.

Similar to the implementation of Ohmic dissipation, we add a constraint on the time step such that (cf. \citealt{Massonetal2012}):
\begin{align}
\label{eq:dtAD}
\Delta t \leq \Delta t_{\rm AD} =&~ C_{\rm AD} \min \left(\frac{\gamma_{\rm AD} \rho_i}{v_A^2} (\Delta x)^2 \right)\nonumber \\
=&~ C_{\rm AD} \min \left(\frac{\gamma_{\rm AD} \rho_i \rho}{B^2} (\Delta x)^2 \right),
\end{align}
with $C_{\rm AD}$ again conservatively chosen to be $0.1$.

\subsection{Particle trajectory integration}
\label{sub:particles}
Particles in DISPATCH can be treated as both massive and massless.  Massless particles may be used for diagnostic tracing of Lagrangian evolution histories, while particles with mass may be used to model the interaction of gas and dust, which is an important process, particularly in protoplanetary disks. 

To be able to represent a wide distribution of dust particle sizes without major impact on the cost the particle solver must be fast, and the particle integration methods have been chosen with that in mind. Trajectories are integrated using a symplectic, kick-drift-kick leap-frog integrator, similar to the one in GADGET2 \citep{springel2005_gadget2}. In the case of massive particles we use forward-time-centring of the drag force, which maintains stability in the stiff limit when drag dominates in the equation of motion. Back-reaction of the drag force on the gas is currently neglected -- this will be included in a future version.

\subsection{Equation-of-state and opacity tables}
\label{sub:microphysics}
DISPATCH is designed from the ground up to include several options and look-up tables for both idealised and realistic equations-of-state (EOS) and opacities. To look-up tables as efficient as possible, binary data files with fine grained tables are generated ahead of time using utilities written in Python and included with the code. These utilities read data from various sources and convert it to look-up tables in the logarithm of density and an energy variable; which type energy is used depends on the solver. For example, the HLLC and ZEUS-3D solvers use thermal energy in the EOS, as do most of the Stagger Code solvers, while one version of the Stagger Code  instead uses entropy as the `energy' variable. Naturally, for simple EOSs such as constant gamma ideal gas, a look-up table is not necessary and calls to the EOS component instead end in a function appropriate for the current solver.

The table look-up is heavily optimised, first by using spline interpolations in the Python utilities to produce tables with sufficient resolution to enable bi-linear interpolation in log-space in DISPATCH. The table address calculations and interpolations are vectorised, while a small fraction of the procedure (data access) remains non-vectorised because of indirect addressing.

\section{Validation}
\label{sec:examples}
Since the solver components of DISPATCH are taken from existing, well-documented codes, validation in the current context requires only verification that the DISPATCH results agree with those from the standalone codes and, furthermore, verification that the partitioning into patches with independent time steps does not significantly affect the results. In this section, we document validation tests for the main HD and MHD solvers, for the non-ideal MHD extension to these solvers, and for the radiative transfer code component. We concentrate on the validation aspect, mentioning performance and scaling only in passing; more details on these aspects will appear in a subsequent paper (Ramsey et al., in prep.).

For any particular experiment, all solvers use the same `generator' to set the initial conditions, which makes it easy to validate the solvers against each other, and to check functionality and reproducibility after changes to the code. A simple regression testing mechanism has been implemented in the form of a shell script, permitting each experiment to be validated against previous or fiducial results. By also making use of the 'pipelines' feature on \texttt{bitbucket.org}, selected validations are automatically triggered whenever an update is pushed to the code's \texttt{git} repository.

\subsection{Supersonic turbulence}
\label{sub:turbulence}
\begin{figure}
  \includegraphics[width=0.5\columnwidth]{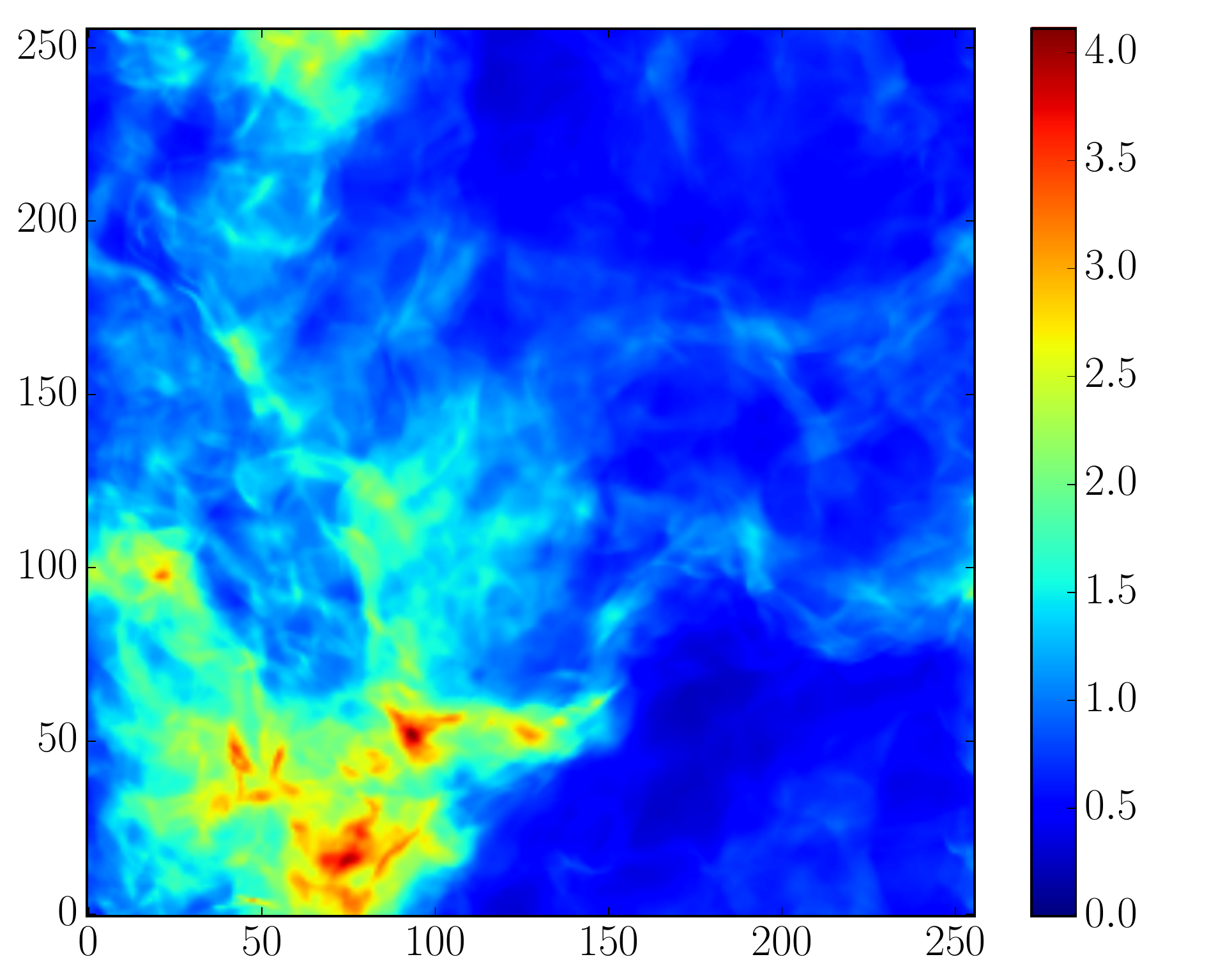}\includegraphics[width=0.5\columnwidth]{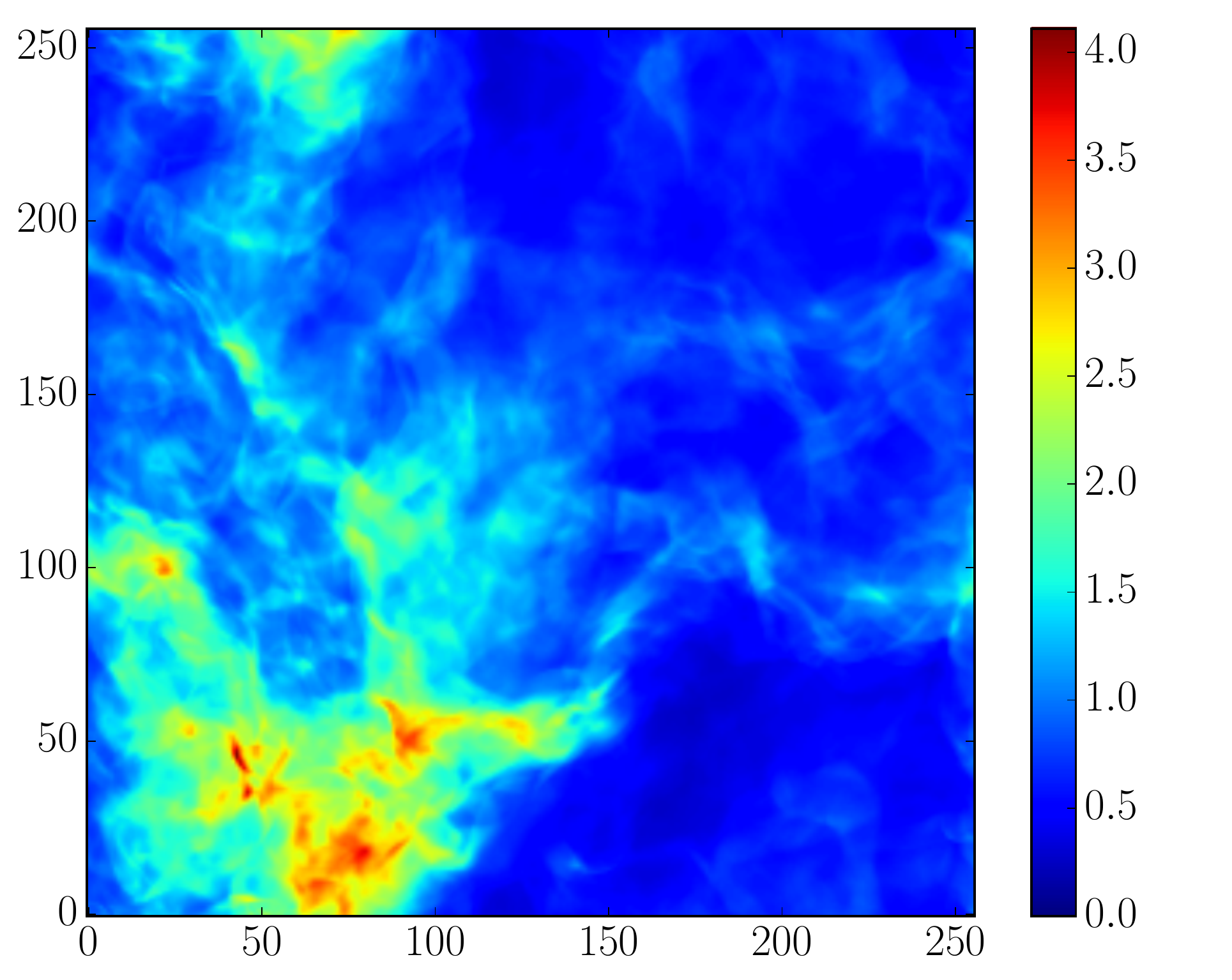}
  \caption{\textit{Left:} Projected density in code units at time $t=0.2$ for the KITP decaying turbulence experiment when using a single patch with $256^3$ cells. \textit{Right:} The same, but using 512 patches, each with $32^3$ cells. Note that he density has been raised to a power of $0.5$ to enhance the contrast. The cell indices are denoted on the axes.}
  \label{fig:turb_decay}
\end{figure}
\begin{figure}
  \includegraphics[width=0.5\columnwidth]{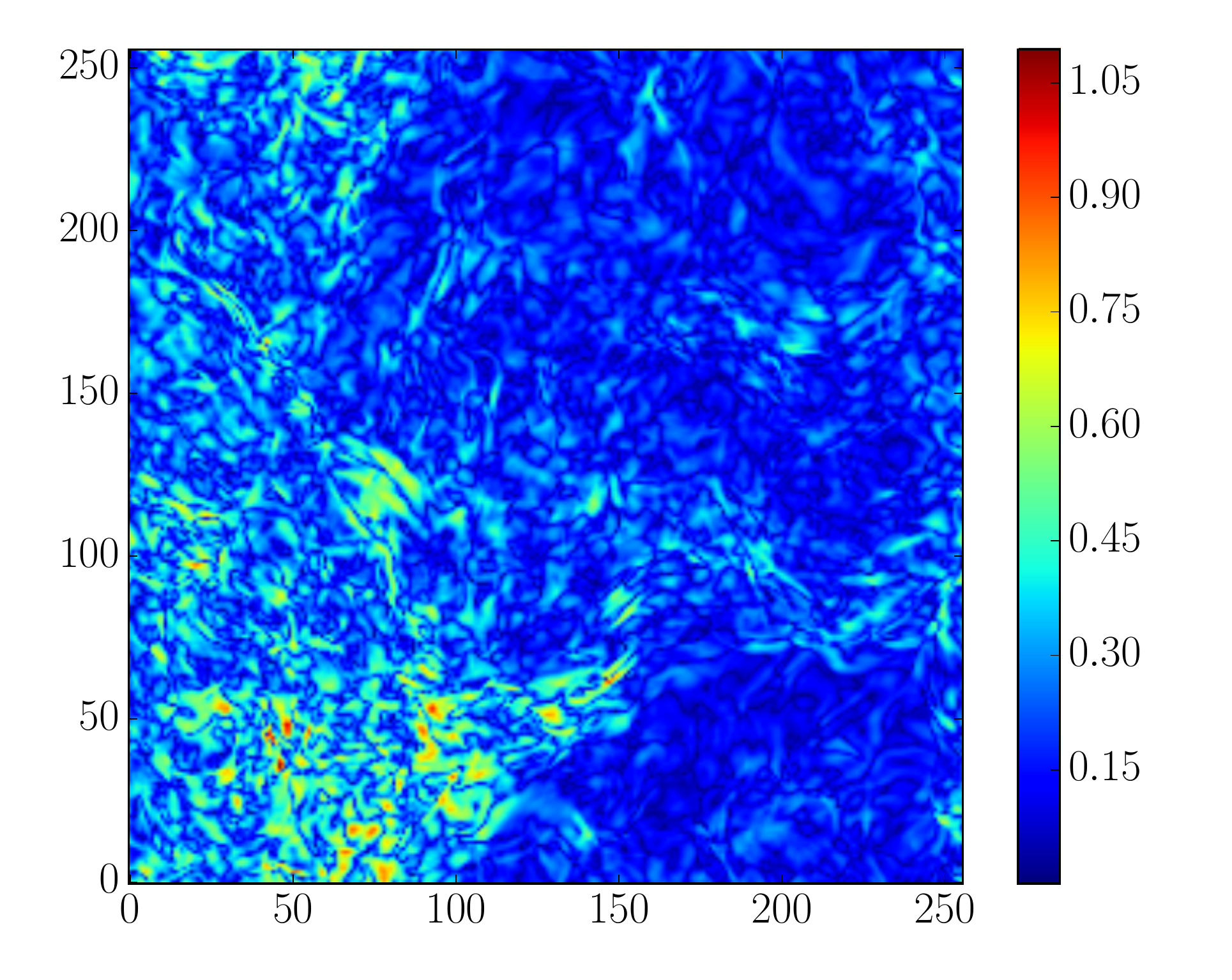}\includegraphics[width=0.5\columnwidth]{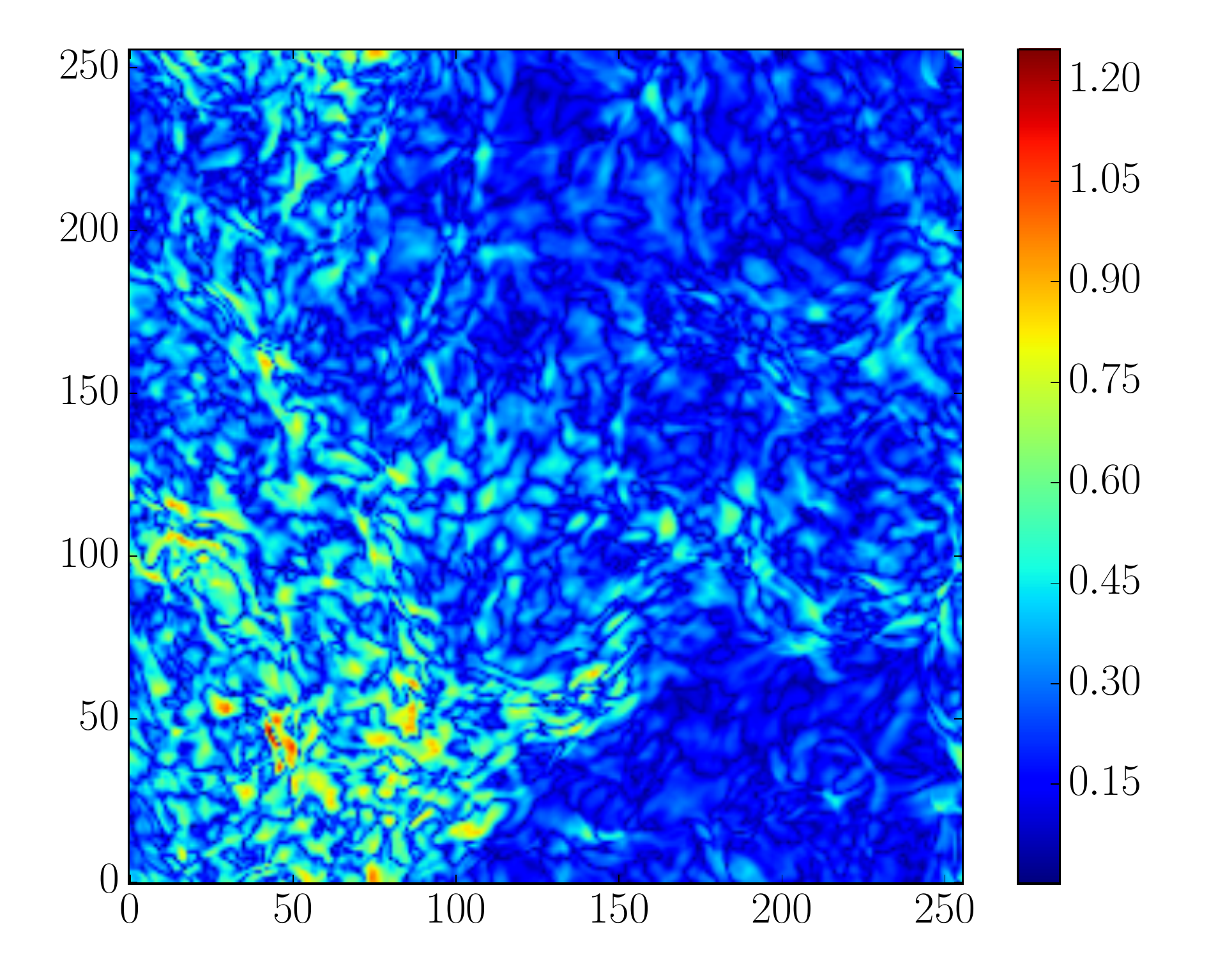}
  \caption{\textit{Left:} The absolute difference between the left and right panels of Figure \ref{fig:turb_decay} (i.e.\ in projected density). \textit{Right:} The difference between two runs, obtained with Courant numbers 0.6 and 0.8, respectively.}
  \label{fig:turb_diff}
\end{figure}

For the first demonstration that the solvers implemented in the DISPATCH framework reproduce results from the stand-alone versions of existing codes, we use the decaying turbulence experiment from the KITP code comparison \citep{kritsuketal2011}. The experiment follows the decay of supersonic turbulence, starting from an archived snapshot with $256^3$ cells at time $t=0.02$, until time $t=0.2$.

In Figure \ref{fig:turb_decay}, we display the projected mass density at the end of the experiment, carried out using the HLLC solver from RAMSES. On the one hand, a single patch (left panel) reproduces results obtained with RAMSES, while, on the other hand, using 512 (right panel) patches with $32^3$ cells each, evolving with local time steps determined by the same Courant conditions as in the RAMSES code (but locally, for each patch), still demonstrates good agreement with the RAMSES results. The solutions are visually nearly indistinguishable, even after roughly two dynamical times. The root-mean-square velocities are practically identical, with a relative difference of only $6\,10^{-4}$ at the end of the experiments.

Figure \ref{fig:turb_diff} shows two difference images, displaying the absolute value of the differences in the projected density (raised to the power of 0.5 to enhance visibility). The left panel shows the difference between the runs with 1 and 512 patches, while the right panel shows the difference between two otherwise identical cases with 512 patches, but with slightly different Courant numbers (0.6 and 0.8). This illustrates that a turbulence experiment can never be reproduced identically, except when all conditions are exactly the same; small differences due to truncation errors grow with time, according to some Lyapunov exponent.

Figures \ref{fig:turb_decay} and \ref{fig:turb_diff} validate the use of our patch-based local time-stepping in two different ways: First, through the absence of any trace of patch boundaries in the difference image and, second, by demonstrating that the differences are consistent with differences expected simply from the use of different Courant numbers. In the single patch case (as well as in RAMSES and other traditional codes), local regions with relatively low flow speeds are effectively evolved with very low local Courant numbers (i.e., time steps are locally much smaller than that permitted by the local Courant condition), while the patch-based evolution takes advantage of the lower speeds by using locally larger time steps.

The local time-step advantage can be quantified by the number of cell-updates needed. The single patch case needs about $2\, 10^{10}$ cell updates (as does RAMSES), while the experiment using 512 patches with local time steps needs only about $1.4\,10^{10}$ cell-updates. Using 20-core Intel Xeon Ivy Bridge CPUs, the 512 patch experiment takes only 6.5 minutes when run on 30 hardware threads (and 7.0 minutes with 20 threads). The cost per cell update is 0.61 core microseconds in the single patch case, 0.56 core-microseconds in the 20 thread case, and 0.53 core-microseconds in the 30 thread case. The super-linear scaling from 1 to 20 cores is mainly due to better cache utilisation when using $32^3$ cells per patch, while the additional improvement when using 30 hardware threads on 20 cores illustrates a hyper-threading advantage; with more hardware threads than cores, the CPU has a better chance to find executable threads.

%%%%%%%%%%%%%%%%%%%%%%%%%%%%%%%%%%%%%%%%%%%%%%%%%%%%%%%%%
\subsection{The local time-step advantage}
\label{sub:localtimestep}

In this benchmark, where the initially highly supersonic turbulence rapidly decays, and a uniform resolution is used, the local time-step advantage is only on the order of 30\%. However, the advantage grows with the complexity of the simulation and, in particular, with the dynamic range of signal speeds present. Two examples, using snapshots from published works may be used to illustrate this point: In zoom-in simulations with RAMSES, such as \citet{kuffmeieretal2016_al26,kuffmeier+17}, time-steps are a factor of two smaller in octs with half the cell size. This is already a significant improvement relative to using the same time step at all AMR resolution levels. However, the global time-step is still set by the highest signal speed that occurs in the domain. In this particular case, this is typically---often not even at the highest resolution level---where the density is very low, but the magnetic field is not. By analysing snapshots from such simulations we find that the update cost per unit time can be reduced by factors of order 10-30 if time-steps are determined locally in patches with dimensions on the order of $32^3$ cells.

Similar situations arise when modelling solar active regions \citep{stein+nordlundAR2012}. The Alfv{\'e}n speed can become extremely large (on the order of 30\% the speed of light) at some height above sunspots where the temperature is low and the density drops much more rapidly with height than the magnetic field. The fractional volume where this happens is very small, but in current simulations the time-step is imposed on the entire computational volume when, in fact, signal speeds are typically only 10--30 km\,s$^{-1}$. Analysis of such snapshots show cost reductions on the order of 30 from using local time-steps. Clearly, the larger the computational volume, the larger the chance to encounter `hotspots' with locally very high signal speeds, and hence the time-step advantage may be expected to become even larger in the future, particularly as models become more realistic and highly resolved.

%%%%%%%%%%%%%%%%%%%%%%%%%%%%%%%%%%%%%%%%%%%%%%%%%%%%%%%%%
\subsection{Weak and strong scaling}
\label{subsub:scaling}
\begin{figure}
  \includegraphics[width=1.0\columnwidth]{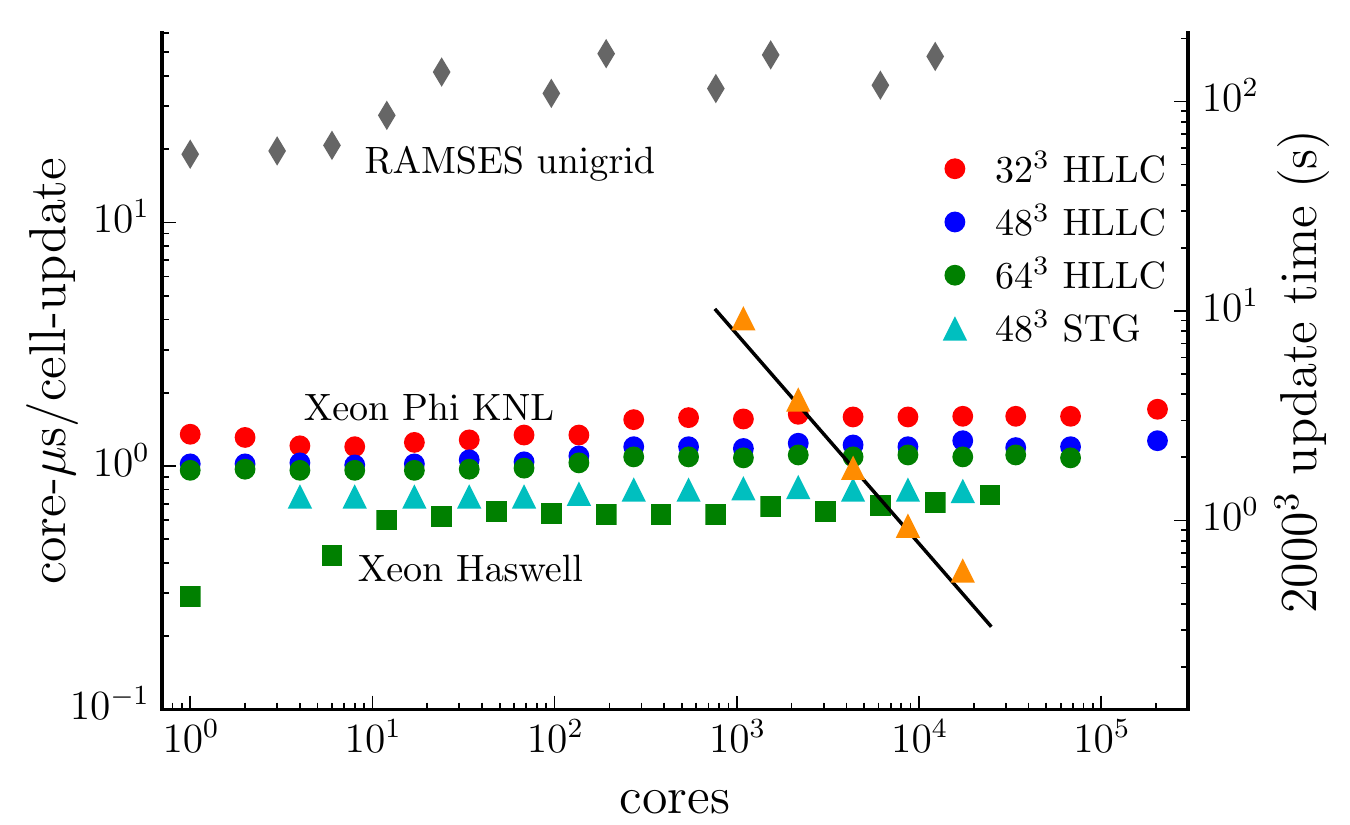}
  \caption{Strong and weak scaling for DISPATCH with the RAMSES HLLC and STAGGER ("STG") code solvers, as applied to a driven turbulence experiment. The performance of the framework is demonstrated using the cost to update one cell as a function of the number of cores. The bottom set of squares denote scaling on Intel Xeon Haswell nodes at HLRS Stuttgart/HazelHen, while the top set of circles (HLLC) and triangles (STAGGER) denote scaling on Intel Xeon Phi Knights Landing (KNL) nodes at CINECA/Marconi (in both cases, with generally two MPI ranks per node). The grey diamonds denote the performance of RAMSES (HLLD) in uniform resolution mode on Intel Xeon Haswell nodes for the same driven turbulence setup. The orange triangles shows strong scaling with 16, 32, 64, 128, and 256 KNL nodes on a $2048^3$ fixed size case, using the STAGGER solver with $32^3$ patches; the straight line indicates ideal strong scaling.}
  \label{fig:scaling}
\end{figure}

\begin{figure}
  \includegraphics[width=0.97\columnwidth]{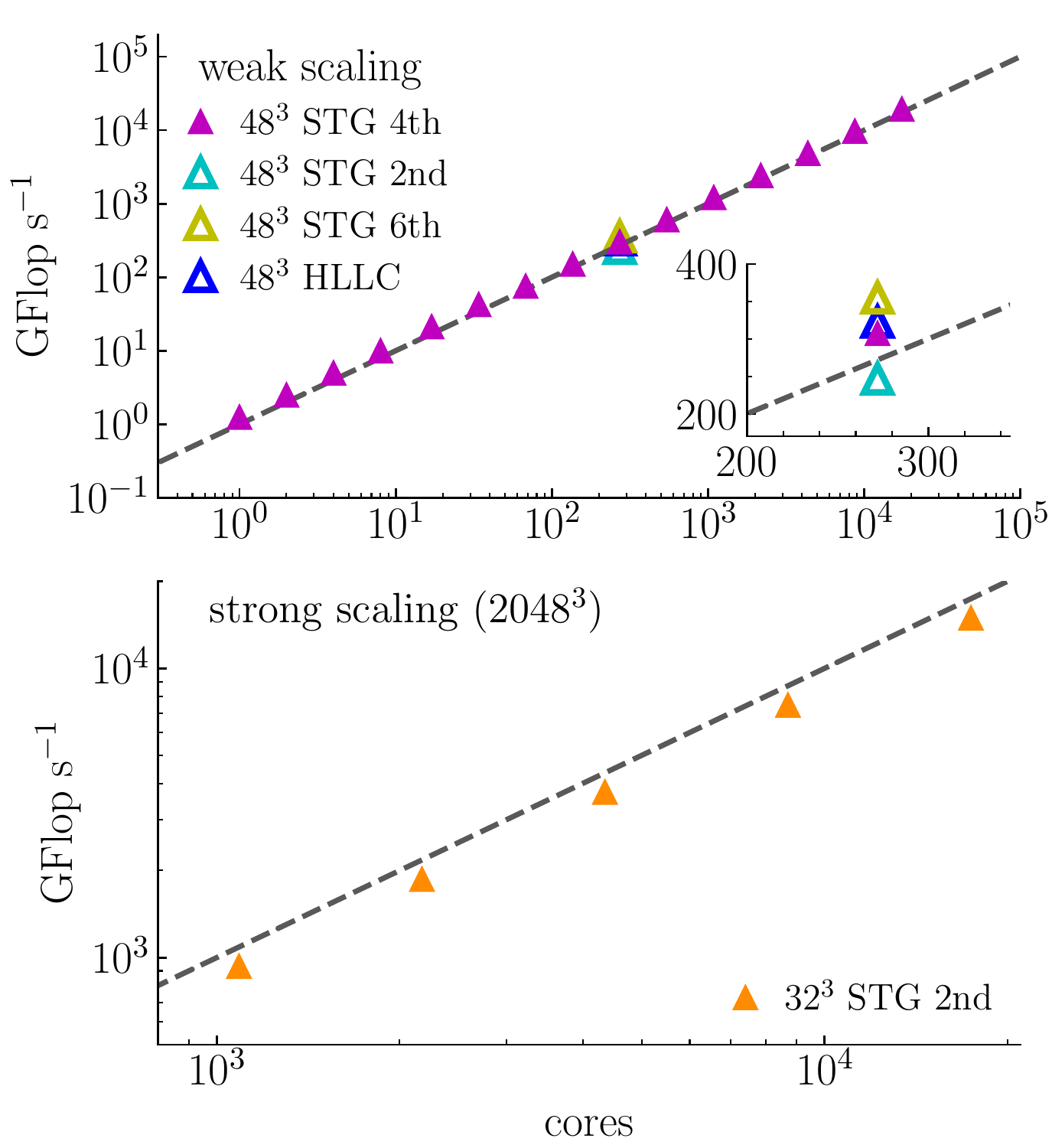}
  \caption{Weak (\textit{top}) and strong (\textit{bottom}) scaling for the driven turbulence experiment expressed in units of GFlop s$^{-1}$ as a function of the number of cores. Dashed lines indicate ideal scaling. The inset zooms in on the differences in GFlop s$^{-1}$ between solvers. All data was collected on CINECA/Marconi. Weak scaling tests were conducted with $48^3$ cells per patch. The strong scaling was conducted with a fixed total $2048^{3}$ cells and $32^3$ cells per patch.}
  \label{fig:flops}
\end{figure}

We now switch to driven isothermal turbulence experiments to explore the weak and strong scaling properties of DISPATCH; the decaying turbulence experiment relies on the $256^3$ code comparison snapshot, and is therefore not suitable for weak scaling tests. In brief, we drive purely solenoidal, supersonic turbulence in a box with wave numbers $kL/2\pi \leq 3.2$, where $L = 1$ is the box size, until $t = 0.2$. The density is initially uniform and equal to 1.0, while the velocities are initially zero. There are nominally 512 patches per MPI rank, each with $32^3-64^3$ cells per patch. The optimal number of cells per patch is problem and hardware dependent; a substantially reduced number of cells per patch results in a degradation in performance (cf.\ \citealt{wombat_2017}), but also allows a potentially larger local time-step advantage.

On Intel Xeon (Haswell) CPUs, we find that, as in the decaying turbulence case, the typical cell-update cost on a single node is on the order of 0.5 -- 0.6 core-microseconds, while on the somewhat slower Intel Xeon Phi (Knights Landing; KNL) cores, the update cost is approximately 0.8 -- 1.1 core-microseconds. As with traditional Xeon CPUs, we find that using $2\times$ hyper-threading on Xeon Phi CPUs reduces the run time by $\sim\! 20\%$. Figure \ref{fig:scaling} demonstrates that DISPATCH has excellent strong scaling properties, subject to the conditions that there needs to be enough work to keep the threads busy, and sufficient memory to hold the time slices. Figure \ref{fig:scaling} also demonstrates that, even with MPI overhead (visible at higher core counts), DISPATCH weak scaling on Intel Xeon and Xeon Phi CPUs is virtually flat up to 1024 (corresponding to 24,576 cores) and 3000 MPI ranks (corresponding to 204,000 cores), respectively. For comparison, we also include performance measurements for RAMSES (using the HLLD solver) operating in uniform resolution mode for the same driven turbulence experiment. After adjusting for the cost difference between HLLD and HLLC solvers (the former is $\sim$5$\times$ more expensive than the latter), evidently, HLLD in DISPATCH is up to 16$\times$ faster than in RAMSES at large core counts. In the RAMSES simulations in, e.g., \citet{padoan+2016,padoan+2017}, the combined effects of HLLD speed, AMR overhead, and non-ideal OpenMP and MPI scaling result in an update cost often exceeding 100 core-microseconds per cell update, indicating that, when combined with expected gains from the local time-step advantage, accrued gain factors on the order of 1000 are possible with DISPATCH in extreme settings.

Figure \ref{fig:flops} presents another metric of the performance of the DISPATCH framework and integrated solvers: the gigaflops per second (GFlop s$^{-1}$) as a function of the number of cores for the driven turbulence experiment. The theoretical single precision peak performance of one Intel Xeon Phi 7250 (KNL) CPU (i.e.\ the ones that constitute CINECA/Marconi-KNL) is $\sim\! 78$ GFlop s$^{-1}$ per core.  The solvers currently implemented in DISPATCH only reach $\sim\! 2 - 3\%$ of this theoretical peak. This is not uncommon for astrophysical fluid codes, which are typically memory bandwidth limited. Note that these values characterize the solvers even in single core execution; the DISPATCH framework controls the scaling of the performance, while the performance per core is determined by the particular solver.

However, while the time per cell update and its scalings (Figure \ref{fig:scaling}) are the main factors determining the `time-to-solution' for a simulation, the number of Flops used per second remains an important complementary indicator; not only is it the measure that \textit{defines} exa-scale, its consideration is also important when trying to minimising the time-to-solution. Modern CPUs can do many Flops per memory load/store, and for essentially all (M)HD solvers, the Flops per second is limited by memory bandwidth rather than by theoretical peak performance. Added complexity per update step thus potentially carries very low extra cost, especially if it does not incur any extra loads/stores. We illustrate this in Figure \ref{fig:flops} by showing the number of GFlop s$^{-1}$ when employing 2nd, 4th, and 6th order operators in the Stagger solver. Although the 4th order operators use five Flops per point, vis-a-vis two Flops per point for the 2nd order operators, the total number of Flops per cell update increases only by a factor of $\sim\! 1.3$ (due to Flops not related to differential operators). Meanwhile, the time per cell update/number of cell updates per core-second remains essentially unchanged, consistent with memory bandwidth being the bottleneck. Since higher order operators reduce the numerical dispersion of the scheme, they potentially allow for a reduction in the number of cells at constant solution quality, and since, in 3-D, the time-to-solution generally scales as the 4th power of the number of cells per dimension, using 4th order operators is a virtually certain advantage. Using 6th order operators further increases the number of flops per core-second while affecting the time per cell update only marginally, potentially offering yet more advantage. However, the increased number of guard zones necessary for higher order operators is another factor that enters the cost balance analysis, and the most optimal choice is thus problem dependent. In current (and future) applications of the framework, we (will) base the choice of solver and order of operators on the quality of the, and the time to, solution using pilot simulations to compare outcomes and to choose measures of `quality' most relevant for each application.

\begin{figure*}
  \includegraphics[width=0.75\textwidth]{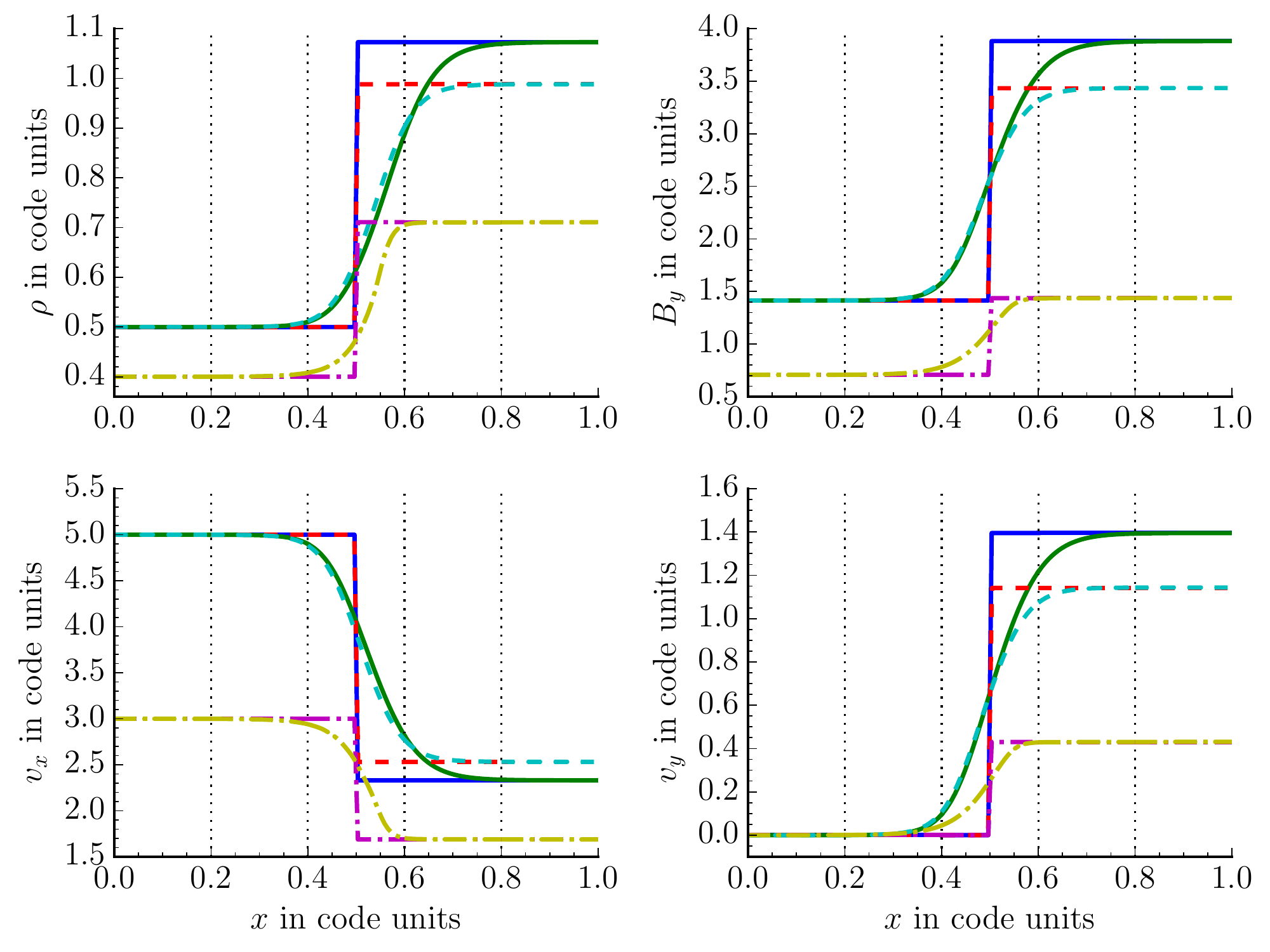}
  \caption{Density (upper left), $B_y$ (upper right), $v_x$ (lower left) and $v_y$ (lower right) for the C-shock experiments. The solid lines correspond to the initial (blue) and evolved (green) states of the isothermal run with AD, the dashed lines correspond to the initial (red) and evolved (cyan) states of the non-isothermal run with AD, the dash-dotted lines correspond to the initial (magenta) and evolved (yellow) states of the non-isothermal run with Ohmic dissipation. Vertical dotted lines denote patch boundaries.}  
  \label{fig:C-shock}
\end{figure*}

\begin{table*}
\centering
{%
\caption{Initial conditions for the non-ideal C-shock experiments.  Pre-shock (upstream) values refer to the left hand side of panels in Figure \ref{fig:C-shock}.}
\begin{tabular}{lrrrrrr}
Variable      & $\rho$  & $v_x$  & $v_y$  & $B_x$        & $B_y$        & $P$    \\
\hline
\hline
Pre-shock  (AD)         & 0.5     & 5      & 0      & $\sqrt[]{2}$ & $\sqrt[]{2}$ & 0.125  \\
Post-shock (AD; isothermal)     & 1.0727  & 2.3305 & 1.3953 & $\sqrt[]{2}$ & 3.8809       & 0.2681 \\
Post-shock (AD; non-isothermal)      & 0.9880  & 2.5303 & 1.1415 & $\sqrt[]{2}$ & 3.4327       & 1.4075 \\
\hline
Pre-shock (Ohm)    & 0.4     & 3      & 0      & $\frac{1}{\sqrt[]{2}}$ & $\frac{1}{\sqrt[]{2}}$ & 0.4 \\
Post-shock (Ohm)   & 0.71084 & 1.68814& 0.4299 & $\frac{1}{\sqrt[]{2}}$ & 1.43667 & 1.19222 \\
\end{tabular} }
\label{tab:C-shock}
\end{table*}

%%%%%%%%%%%%%%%%%%%%%%%%%%%%%%%%%%%%%%%%%%%%%%%%%%%%%%%%%
\subsection{Non-ideal MHD: C-shock}
\label{sub:cshock}

To validate the non-ideal MHD extensions, we carry out C-shock experiments following \citet{Massonetal2012} for both Ohmic dissipation and ambipolar diffusion. Table \ref{tab:C-shock} shows the initial (left and right) states. In the isothermal case, we set the adiabatic index $\gamma = 1+10^{-7}$, while we use $\gamma = 5/3$ for the non-isothermal and Ohmic dissipation runs. In the ambipolar diffusion runs, we set $\gamma_{\rm AD} = 75$, and in the Ohmic dissipation run we use $\eta_{\rm Ohm} = 0.1$, where we assumed an ion density of $\rho_i = 1$ in all three cases. We test the implementations using five patches, each with dimensions 30x1x1, and plot the results in Figure \ref{fig:C-shock}. Our results are in good agreement with \citet{Massonetal2012}. Like \citet{Massonetal2012} and \citet{duffinetal2008}, we find that the shock undergoes a brief period of adjustment before reaching a steady state solution in all cases. We have confirmed that, for a moving C-shock, the solution quality is not affected by crossing patch boundaries.

Even for a simple benchmark like the non-ideal C-shock, the DISPATCH local time-stepping advantage is already beginning to show: In the ambipolar diffusion experiments, there is a factor of $\sim\! 2$ difference in the time steps between patches\footnote{In the Ohmic dissipation experiment, since the Ohmic time step depends only on (a constant) $\eta_{\rm Ohm}$ and the grid spacing, the time step is the same in every patch.}. As the complexity of the problem and the number of patches increases, the advantage for non-ideal MHD experiments is expected to increase. Indeed, we are currently exploring if DISPATCH can compete with super-time-stepping algorithms (e.g.\ \citealt{alexiadesetal1996_sts,meyeretal2014_sts}) for ambipolar diffusion, or even if there is a greater benefit by combining them.

\subsection{Radiative transfer: Shadow casting benchmark}
\label{sub:rtshadow}
\begin{figure}
  \includegraphics[width=1.0\columnwidth]{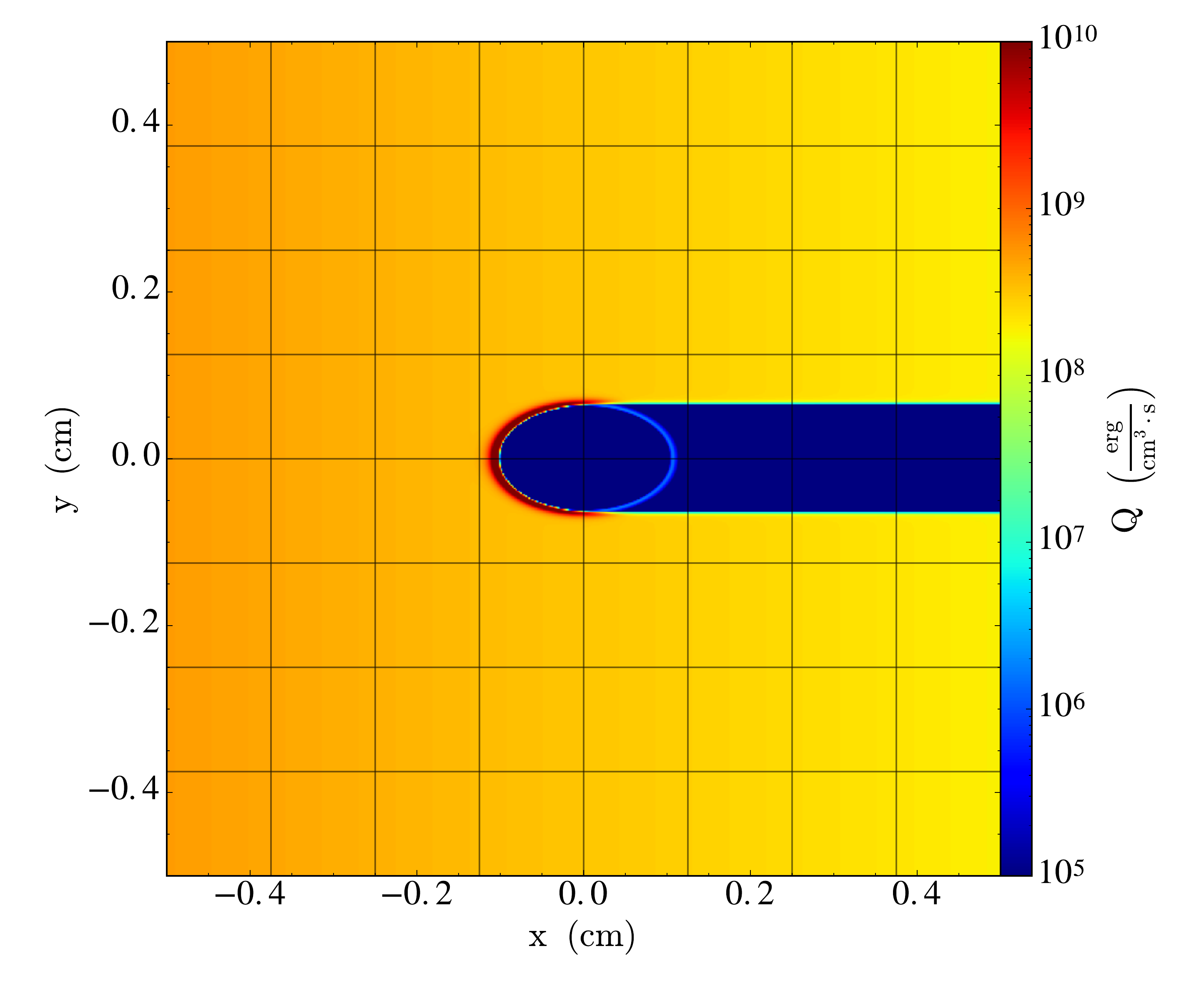}
  \caption{Results for the shadow casting benchmark. The irradiation source is located at the left boundary. The heating rate per unit volume, $Q$ (Eq.\ \ref{eq:heating}), is shown, with levels indicated by the colour bar to the right.}
  \label{fig:rt_shadow}
\end{figure}

To test our radiative transfer implementation, we apply the so-called `shadow' casting benchmark (e.g.\ \citealt{hayesnorman2003, Jiang2012, ramseydullemond15}). We consider a 3-D Cartesian domain of size $-0.5 \leq [x,y,z] \leq 0.5$ cm. An overdense ellipsoid is placed at the origin, with a density profile prescribed by:
\begin{equation}
\label{eq:shadow_rho}
\rho (x,y,z) = \rho_0 + \frac {\rho_1 - \rho_0} {1+ e^{10(r-1)}},
\end{equation}
where $r=(x/a)^2 + (y/b)^2 + (z/b)^2$, $a= 0.10$ cm, $b=0.06$ cm, and $\rho_1 = 10 \rho_0$. The ambient medium is initialised with a gas temperature $T_0 = 290$ K and a density $\rho_0 = 1$ g cm$^{-3}$. The entire domain is in pressure equilibrium. The gas opacity is given by:
\begin{equation}
\label{eq:shadow_opac}
\alpha = \rho \kappa = \alpha_0 \left(\frac{T}{T_0}\right)^{-3.5}\left(\frac{\rho}{\rho_0} \right)^2,
\end{equation}
where $\alpha_0=1$. We apply a constant and uniform irradiation source across the left boundary (at $x = -0.5$ cm), characterised by blackbody emission with an effective temperature T$_{\rm irr} = $6 T$_{0}$ K. The other boundaries are assumed to be blackbodies with effective temperature T$_0$. For the experiment, we use 8$^3$ patches, each containing 64$^3$cells. The ray geometry contains 13 directions with 45$^{\circ}$ space-angle separation.

\begin{figure*}
  \includegraphics[width=0.33\textwidth]{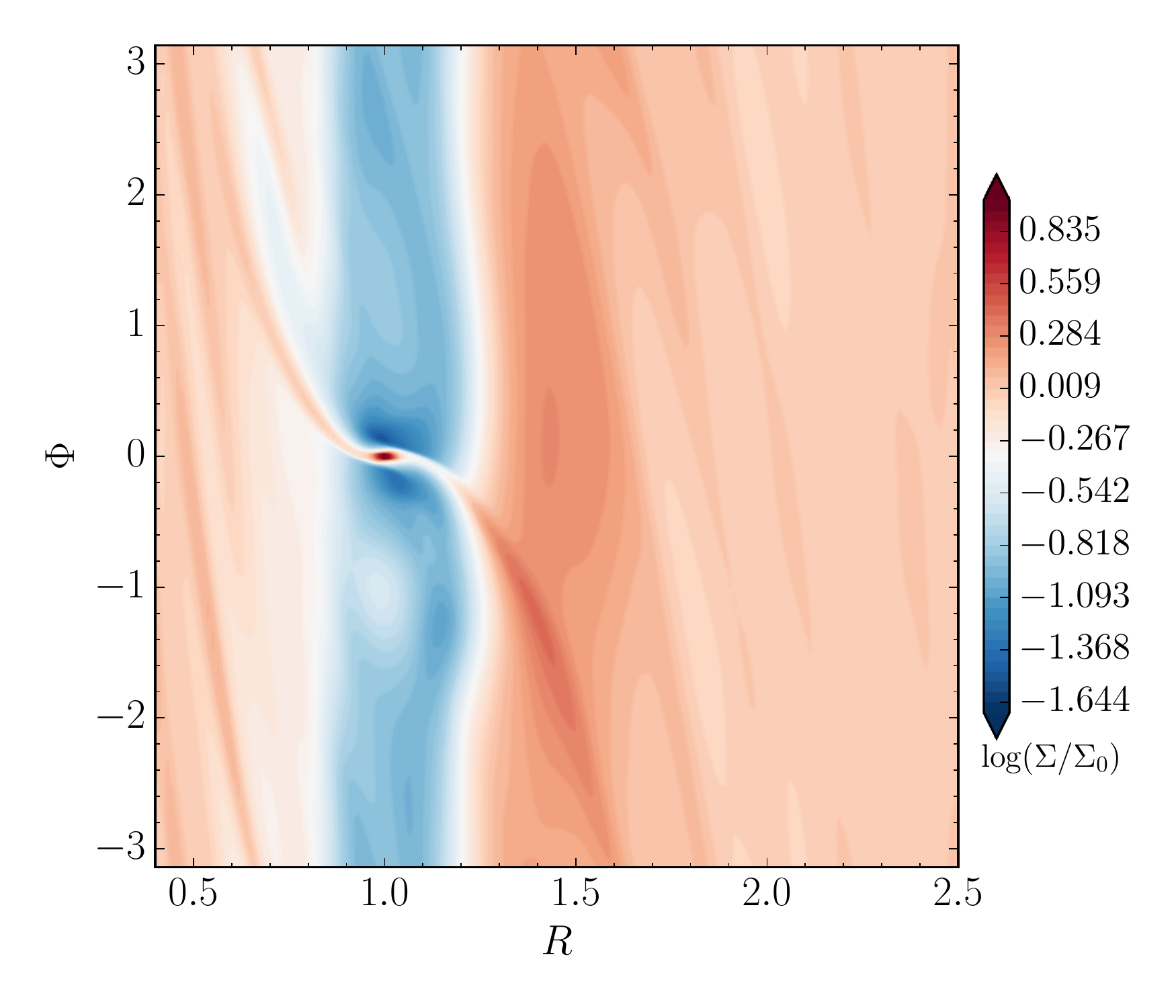}\includegraphics[width=0.33\textwidth]{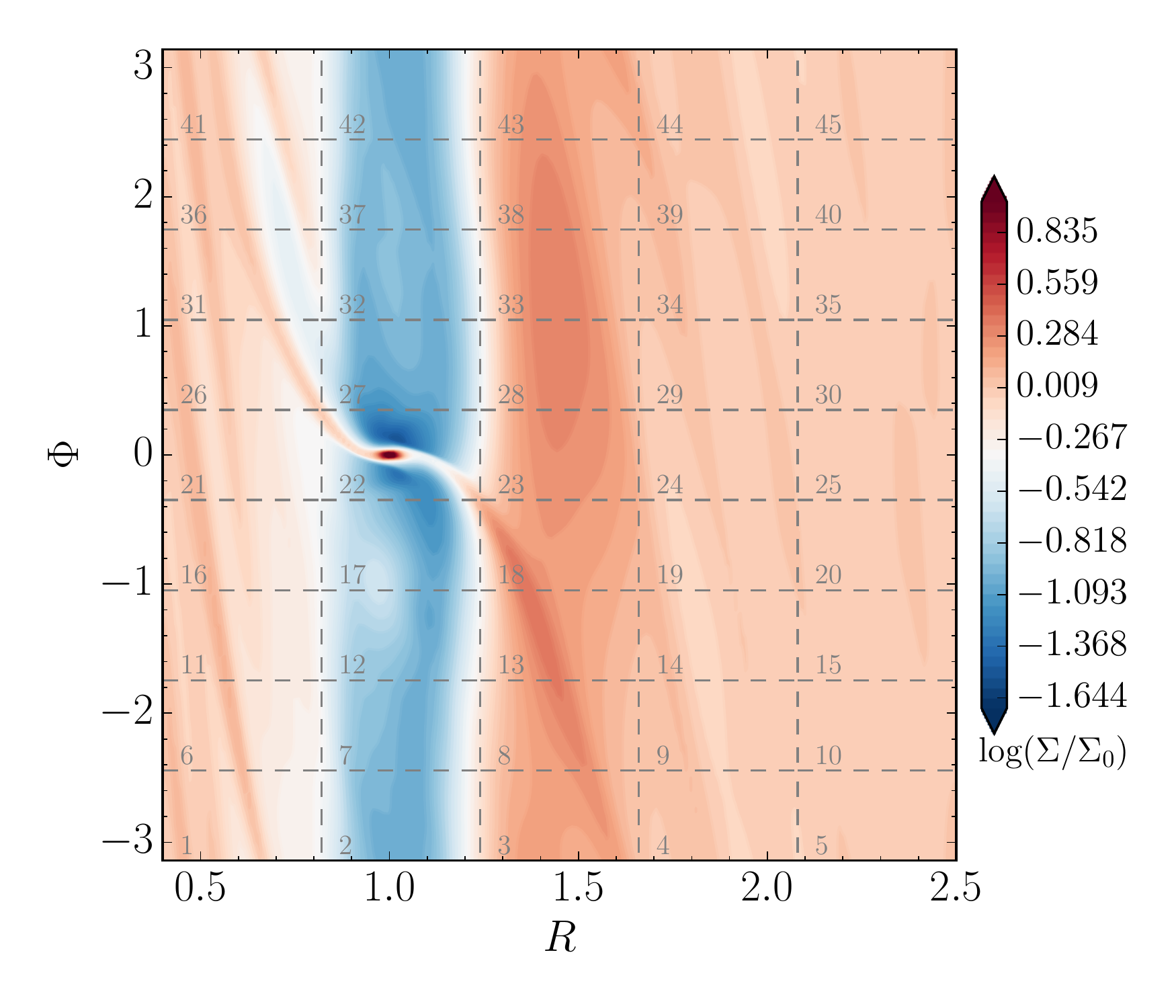}\includegraphics[width=0.33\textwidth]{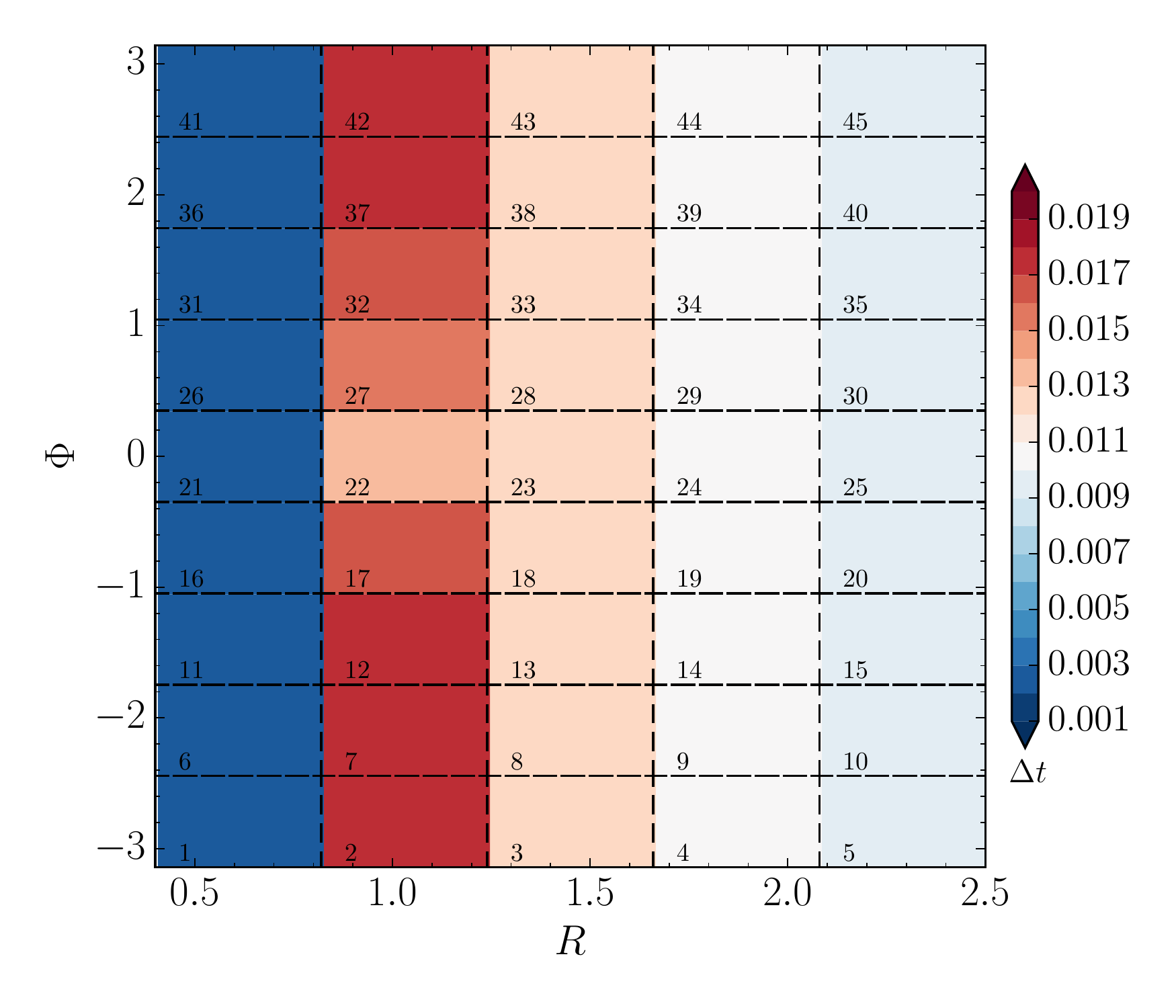}
  \caption{\textit{Left:} A $1M_\mathrm{J}$ planet simulated for 100 orbits using a single patch with $n_R \times n_\Phi = 160 \times 450$ uniformly-spaced zones. \textit{Middle:} The same but subdividing the domain into 45 equally-sized patches. \textit{Right:} The time step in each patch after 100 orbits.}
  \label{fig:devalborro}
\end{figure*}

In this setup, the ambient medium is optically thin and the photon mean free path parallel to a coordinate axis is equal to the length of the domain. The clump is, meanwhile, optically thick and has an interior mean free path of $\approx 3.2\, 10^{-6}$ cm. The clump thus absorbs the incident irradiation, and subsequently casts a shadow behind it. Figure \ref{fig:rt_shadow} shows the resulting heating rate per unit volume, which is given by:
\begin{equation}\label{eq:heating}
Q = \rho\kappa (J - S),
\end{equation}
where $J$ is the mean intensity and $S$ is the source function. The solution displays, qualitatively and quantitatively, all the expected features: the shadow is sharp, it has a finite-size border, the front of the clump is strongly heated due to the incident radiation, and the back side of the clump experiences a very mild heating due to the incident thermal radiation from the ambient medium. The heating rates, meanwhile, compare well with the results of \citet{ramseydullemond15}.

For this experiment, the update cost is $\sim\! 0.6$ core-microseconds per cell on Intel Xeon Ivy Bridge CPUs, including RT. Running the same experiment with RT disabled instead results in a cost of $\sim\! 0.5$ core-microseconds per cell. Thus, even with 13 spatial directions, and even with solving for RT after every MHD update, the RT algorithm only adds approximately 20\% to the experiment cost. Although using a more realistic EOS (i.e.\ look-up table) will increase the cost slightly, clearly, the DISPATCH RT algorithm can provide accurate RT solutions at a very low cost. In a subsequent paper of this series (Popovas et al., in prep.), additional details and benchmarks, including further documentation on the accuracy and speed of RT in DISPATCH, will be provided.

\subsection{de Val-Borro disk benchmark}
\label{sub:devalborro}

One of the current and primary science goals of DISPATCH is to model planet formation processes in global and realistic settings. As such, we now apply DISPATCH to the global disk benchmark of \citet{devalborroetal2006}. More specifically, we embed a Jupiter-like planet in an initially uniform surface density protoplanetary accretion disk and then allow the disk to evolve for 100 orbits at the semi-major axis of the planet. Following \citet{devalborroetal2006}, we initialise a two-dimensional, geometrically thin disk with a constant surface density given by $\Sigma_0 = 0.002 M_* / \pi a^2$, where $M_*$ is the mass of the central star and $a$ is the semi-major axis of the planet. We use an aspect ratio of $H / R = 0.05$, where $H$ is the canonical disk scale height, an effectively isothermal EOS ($\gamma = 1.0001$), and an initially Keplerian orbital velocity. 

In scaled units, the mass of the central star is set to $1.0$, while the planet is assigned a mass of $10^{-3}$ and a semi-major axis of $1.0$. The simulation is performed in the reference frame co-rotating with the planet and centred on the central star. Therefore, the gravitational potential of the system is comprised of one term each for the star and planet, plus an additional `indirect' term to account for the fact that the centre-of-mass of the system does not correspond to the origin (cf.\ \citealt{dangelobodenheimer2013}):
\begin{equation}
  \Phi = -\frac{GM_*}{R} - \frac{GM_\mathrm{p}}{\sqrt{|\mathbfit{R} - \mathbfit{R}_\mathrm{p}|^2 + \epsilon^2}} + \frac{GM_\mathrm{p}}{R^3_\mathrm{p}}\mathbfit{R}\cdot\mathbfit{R}_\mathrm{p},
  \label{eq:devalborro_gravpot}
\end{equation}
where $G$ is the gravitational constant, $M_\mathrm{p}$ is the mass of the planet, $R$ is the cylindrical radius, $R_\mathrm{p}$ is the position of the planet in cylindrical coordinates, and $\epsilon$ is the gravitational softening length. Still following \citet{devalborroetal2006}, we set $\epsilon = 0.6 H$, and a planetary position of $(Z,R,\Phi) = (0.0,1.0,0.0)$. In the co-rotating frame, the planet's position is fixed, i.e., the planet is not permitted to migrate.

The computational domain has dimensions $[0.4a, 2.5a]$ in radius and $[-\pi, \pi]$ in azimuth; we use polar cylindrical coordinates and the ZEUS-3D solver. When using curvilinear coordinates with ZEUS-3D, it is the angular rather than the linear momentum that is solved, thus guaranteeing the conservation of angular momentum \citep{kley1998}. Code parameters of note are the Courant factor, which is set to $0.5$, and the artificial viscosity parameters \citep{clarke2010}, \verb|qcon| and \verb|qlin|, are set to $2.0$ and $0.1$, respectively. In the results presented below, a kinematic viscosity of $10^{-5}$ (in units where $GM_*$ = 1 and $a = 1$) is used. 

We adopt periodic boundary conditions in the azimuthal direction, and the `wave-killing' conditions of \citet{devalborroetal2006} in the radial direction, with damping activated in the ranges $[0.4a, 0.5a]$ and $[2.1a, 2.5a]$. It is worth noting that we do not gradually grow the planet mass over the first few orbits (cf.\ \citealt{devalborroetal2006}; Sect.\ 3.1), but instead begin with the full planetary mass in place.

The left panel of Figure \ref{fig:devalborro} shows the benchmark results for a single patch at a resolution of $n_R \times n_\Phi = 160 \times 450$ cells after 100 orbits. These results can be directly and favourably compared with Figure 10 of \citet{devalborroetal2006}. The middle panel, meanwhile, shows the results from sub-dividing the domain into 45 equally-sized patches. As can be seen, the model with 45 individual patches is nearly identical to the model with a single patch. Finally, the right panel illustrates the time step in each patch after 100 orbits; there is a factor of $\sim\!8.2$ difference between the largest and smallest time steps. This local time-stepping advantage, over the course of 100 orbits, reduces the CPU time required by the multiple-patch version by a factor of roughly two relative to the single-patch version ($\sim\! 6$ hr vs.\ $\sim\! 12$ hr).

%%%%%%%%%%%%%%%%%%%%%%%%%%%%%%%%%%%%%%%%%%%%%%%%%
\section{Summary and outlook}
\label{sec:summary}
We have introduced a hybrid MPI/OpenMP code framework that permits semi-independent task-based execution of code, e.g., for the purpose of updating a collection of semi-independent patches in space-time. OpenMP parallelism is handled entirely by the code framework -- already existing code need not (and should not) contain OpenMP parallel constructs beyond declaring appropriate variables as \verb|threadprivate|. OpenMP tasks are generated by a rank-local \textit{dispatcher}, which also selects the tasks that are ready for (and most in need of) updating. This typically implies having valid guard zone data imported from neighbours, but can take the form of any type of inter-task dependency. The type of tasks can vary, with HD, ideal MHD, radiative transfer and non-ideal MHD demonstrated here; particle-in-cell (PIC) methods coupled to MHD methods on neighbouring patches are currently in development. Tasks do not necessarily have to be grid-based (allowing,, e.g., particle-based tasks), while tasks that are may use either Cartesian or curvilinear meshes. The code framework supports both stationary and moving patches, as well as both static and dynamic mesh refinement; the implementation details for moving patches and dynamic refinement will be featured in a forthcoming paper (Ramsey et al., in prep.).

As DISPATCH is targeting exa-scale computing, a feature of great importance for the performance of the framework is that time steps are determined by local conditions (e.g.\ the Courant condition in the case of mesh-based tasks/patches); this can lead to potentially significant reductions in computing time in cases where the signal speed varies greatly within the computational domain. Patches, in particular, are surrounded by a sufficient number of guard zones to allow interior cells to be updated independently of other patches within a single time step; the guard zone data is retrieved or interpolated from neighbouring patches. Since patches evolve with different time steps, it is necessary to save a number of time slices for interpolation (and possibly extrapolation) to the time whence guard zone values are required. Performance, even for existing codes, improves when running under the DISPATCH framework for several reasons: First, patches are small enough to utilise cache memory efficiently, while simultaneously being large enough to obtain good vectorisation efficiency. Secondly, task overloading ensures all OpenMP threads remain constantly busy, which results in nearly linear OpenMP speed-up, even for a very large numbers of cores (e.g.\ Intel Xeon Phi). Third, load imbalance caused by multi-physics modules with significantly different update costs per unit time is automatically compensated for, since the task scheduling automatically favours tasks that tend to lag behind. Fourth, each MPI process (typically one per socket per node) only communicates with its nearest neighbours -- here interpreted broadly as those MPI processes that are either geometrically close, or else causally coupled via, for example, radiative transfer. Finally, load balancing is designed to minimise work imbalance while simultaneously keeping the number of communications links near a minimum. As demonstrated in Figures \ref{fig:scaling} and \ref{fig:flops}, these features result in a code framework with excellent OpenMP and MPI scaling properties. 

In protoplanetary disks, for example, the advantages of DISPATCH relative to a conventional adaptive mesh refinement code such as RAMSES are, first of all, a gain from co-moving patches and, second, an additional advantage due to local time-stepping. Indeed, analysis of snapshots from recent zoom-in simulations performed with RAMSES \citep{nordlund+14, kuffmeieretal2016_al26,kuffmeier+17} demonstrate that the cost reduction from local time-stepping and co-moving patches can be on the order of 10-30 in such simulations. Beyond that, the implementation of the HLLD solver in DISPATCH exhibits a substantial raw speed improvement per cell update by a factor of about 16 relative to the implementation in RAMSES -- in large part due to the much smaller guard zone overhead (the 2x2x2 octs used in RAMSES need 6x6x6 cells for updates), but also due to contributions from better vectorisation and cache utilisation. In extreme cases, the accrued reductions in raw update cost and AMR overhead, improved OpenMP and MPI scaling, plus co-moving patches and local time-steps, can result in cost reductions that reach factors on the order of 1000. Meanwhile, under simpler conditions where the resolution is constant, and the dynamic range of signal speeds is small, the cost reduction will be limited to mainly the difference in raw cell update cost. Relative to AMR codes that work with blocks of dimensions similar to DISPATCH, the advantages lie mainly in the gains from minimal AMR overhead, local time-stepping, and co-moving patches.

In DISPATCH, the number of tasks per MPI rank (typically one per compute node or CPU socket) is kept sufficiently high to ensure all of the cores within a rank stay busy at all times. As the capacity (i.e.\ the number of cores) of compute nodes is growing with time, in the future, each rank will be able to handle more tasks, and hence an increasing fraction of tasks will be `internal', i.e., their neighbour tasks will reside on the same MPI rank, and thus will not require communication with neighbouring ranks. By construction then, as the capacity grows, each node remains fully occupied, and inter-node communication needs decrease rather than increase with growing problem size. Combined with the fact that, in DISPATCH, each rank only communicates with its nearest neighbours, the OpenMP and MPI scaling characteristics of DISPATCH therefore satisfy our definition of being exa-scale ready.

Planned and ongoing applications of the DISPATCH code framework include pebble accretion via hierarchically resolved Hill spheres around planetary embryos \citep{popovasetal2018_arXiv}, solar and stellar magneto-convection with chromospheric and coronal activity, with further extensions applying a combined MHD-PIC multiple-domain-multiple-physics method to particle acceleration in model coronae. Solar and stellar magneto-convection is a context where the local time-step advantage can become very large. Analysis of snapshots from \citet{stein+nordlundAR2012} shows a gain of a factor $\sim 30$, caused by the contrast between the extremely short time steps required in the low density atmosphere above sunspots, and the generally much longer time steps allowed at most other locations. In both this context and the star formation context, the local time-step advantage is expected to become even more significant in the future as models grow in physical size and complexity.

Part of the motivation for developing DISPATCH was to create a framework that is suitable for studying planet formation. Recall, for example, the scene including a protostellar system modelled by using a sink particle component plus MESA stellar evolution model for the central star, plus a collection of moving, orbiting patches to represent the gas in the protostellar accretion disk (Sect.\ \ref{sub:scenes}). Likewise, particle tasks could be operating to represent the dust, either as sub-tasks inside other tasks, or as semi-independent separate tasks. These would then be able to use data on gas and dust properties from accretion disk patches to estimate the production rate of thermally processed components -- representing, for example, chondrules and high-temperature-condensates (calcium-aluminium inclusions, amoeboid olivine aggregates, hibonites, etc.; cf.\ \citealt{haugboelle+17}), and would be able to model the subsequent transport of thermally processed components by disk winds and jet outflows (e.g.\ \citealt{bizzarro+17}).

The DISPATCH framework will be made open-source and publicly available in a not too distant future.

\section*{Acknowledgements}

The work of \AA N, MK, and AP was supported by a grant from the Danish Council for Independent Research (DFF grant 1323-00199B). The Centre for Star and Planet Formation is funded by the Danish National Research Foundation (DNRF97). Storage and computing resources at the University of Copenhagen HPC centre, funded in part by Villum Fonden (VKR023406), were used to carry out some of the simulations presented here. We also acknowledge PRACE for awarding us preparatory access (2010PA3402) to Hazel Hen at HLRS Stuttgart to obtain scaling data. We further acknowledge Klaus Galsgaard for allowing us to obtain KNL scaling data on CINECA/Marconi under his PRACE grant 2016143286 and CINECA support staff for helping us obtain scaling measurements at 204,000 cores.

%%%%%%%%%%%%%%%%%%%%%%%%%%%%%%%%%%%%%%%%%%%%%%%%%%



\clearpage
\bibliographystyle{mnras}
\bibliography{methods} % if your bibtex file is called example.bib

\begin{thebibliography}{}
\makeatletter
\relax
\def\mn@urlcharsother{\let\do\@makeother \do\$\do\&\do\#\do\^\do\_\do\%\do\~}
\def\mn@doi{\begingroup\mn@urlcharsother \@ifnextchar [ {\mn@doi@}
  {\mn@doi@[]}}
\def\mn@doi@[#1]#2{\def\@tempa{#1}\ifx\@tempa\@empty \href
  {http://dx.doi.org/#2} {doi:#2}\else \href {http://dx.doi.org/#2} {#1}\fi
  \endgroup}
\def\mn@eprint#1#2{\mn@eprint@#1:#2::\@nil}
\def\mn@eprint@arXiv#1{\href {http://arxiv.org/abs/#1} {{\tt arXiv:#1}}}
\def\mn@eprint@dblp#1{\href {http://dblp.uni-trier.de/rec/bibtex/#1.xml}
  {dblp:#1}}
\def\mn@eprint@#1:#2:#3:#4\@nil{\def\@tempa {#1}\def\@tempb {#2}\def\@tempc
  {#3}\ifx \@tempc \@empty \let \@tempc \@tempb \let \@tempb \@tempa \fi \ifx
  \@tempb \@empty \def\@tempb {arXiv}\fi \@ifundefined
  {mn@eprint@\@tempb}{\@tempb:\@tempc}{\expandafter \expandafter \csname
  mn@eprint@\@tempb\endcsname \expandafter{\@tempc}}}

\bibitem[\protect\citeauthoryear{Adams et~al.,}{Adams
  et~al.}{2015}]{adamsetal2015_chombo}
Adams M.,  et~al., 2015, Chombo Software Package for AMR Applications - Design
  Document, \url {http://crd.lbl.gov/assets/pubs_presos/chomboDesign.pdf}

\bibitem[\protect\citeauthoryear{Alexiades, Amiez  \& Gremaud}{Alexiades
  et~al.}{1996}]{alexiadesetal1996_sts}
Alexiades V.,  Amiez G.,   Gremaud P.-A.,  1996, \mn@doi [Communications in
  Numerical Methods in Engineering]
  {10.1002/(SICI)1099-0887(199601)12:1<31::AID-CNM950>3.0.CO;2-5}, 12, 31

\bibitem[\protect\citeauthoryear{{Almgren} et~al.,}{{Almgren}
  et~al.}{2010}]{almgrenetal2010_castro}
{Almgren} A.~S.,  et~al., 2010, \mn@doi [\apj] {10.1088/0004-637X/715/2/1221},
  \href {http://adsabs.harvard.edu/abs/2010ApJ...715.1221A} {715, 1221}

\bibitem[\protect\citeauthoryear{{Baumann}, {Haugb{\o}lle}  \&
  {Nordlund}}{{Baumann} et~al.}{2013}]{baumannetal2013}
{Baumann} G.,  {Haugb{\o}lle} T.,   {Nordlund} {\AA}.,  2013, \mn@doi [\apj]
  {10.1088/0004-637X/771/2/93}, \href
  {http://adsabs.harvard.edu/abs/2013ApJ...771...93B} {771, 93}

\bibitem[\protect\citeauthoryear{{Ben{\'{\i}}tez-Llambay} \&
  {Masset}}{{Ben{\'{\i}}tez-Llambay} \&
  {Masset}}{2016}]{benitezllambaymasset2016}
{Ben{\'{\i}}tez-Llambay} P.,  {Masset} F.~S.,  2016, \mn@doi [\apjs]
  {10.3847/0067-0049/223/1/11}, \href
  {http://adsabs.harvard.edu/abs/2016ApJS..223...11B} {223, 11}

\bibitem[\protect\citeauthoryear{{Berger} \& {Colella}}{{Berger} \&
  {Colella}}{1989}]{bc89}
{Berger} M.~J.,  {Colella} P.,  1989, \mn@doi [Journal of Computational
  Physics] {10.1016/0021-9991(89)90035-1}, \href
  {http://adsabs.harvard.edu/abs/1989JCoPh..82...64B} {82, 64}

\bibitem[\protect\citeauthoryear{{Berger} \& {Oliger}}{{Berger} \&
  {Oliger}}{1984}]{bergeroliger84}
{Berger} M.~J.,  {Oliger} J.,  1984, \mn@doi [Journal of Computational Physics]
  {10.1016/0021-9991(84)90073-1}, \href
  {http://adsabs.harvard.edu/abs/1984JCoPh..53..484B} {53, 484}

\bibitem[\protect\citeauthoryear{Berzins, Luitjens, Meng, Harman, Wight  \&
  Peterson}{Berzins et~al.}{2010}]{berzinsetal2010_uintah}
Berzins M.,  Luitjens J.,  Meng Q.,  Harman T.,  Wight C.~A.,   Peterson J.~R.,
   2010, in Proceedings of the 2010 TeraGrid Conference. TG '10.
ACM, New York, NY, USA, pp 3:1--3:8, \mn@doi{10.1145/1838574.1838577}, \url
  {http://doi.acm.org/10.1145/1838574.1838577}

\bibitem[\protect\citeauthoryear{{Bizzarro}, {Wielandt}, {Haugb{\o}lle}  \&
  {Nordlund}}{{Bizzarro} et~al.}{2017}]{bizzarro+17}
{Bizzarro} M.,  {Wielandt} D.,  {Haugb{\o}lle} T.,   {Nordlund} A.,  2017, LPI
  Contributions, \href {http://adsabs.harvard.edu/abs/2017LPICo1975.2008B}
  {1975, 2008}

\bibitem[\protect\citeauthoryear{{Brandenburg} \& {Dobler}}{{Brandenburg} \&
  {Dobler}}{2002}]{brandenburgdobler2002_pencil}
{Brandenburg} A.,  {Dobler} W.,  2002, \mn@doi [Computer Physics
  Communications] {10.1016/S0010-4655(02)00334-X}, \href
  {http://adsabs.harvard.edu/abs/2002CoPhC.147..471B} {147, 471}

\bibitem[\protect\citeauthoryear{Brown, Henshaw  \& Quinlan}{Brown
  et~al.}{1997}]{brownetal1997_overture}
Brown D.~L.,  Henshaw W.~D.,   Quinlan D.~J.,  1997, Overture: An
  object-oriented framework for solving partial differential equations.
Springer Berlin Heidelberg, Berlin, Heidelberg, pp 177--184,
  \mn@doi{10.1007/3-540-63827-X_59}, \url
  {http://dx.doi.org/10.1007/3-540-63827-X_59}

\bibitem[\protect\citeauthoryear{{Bryan} et~al.,}{{Bryan}
  et~al.}{2014}]{bryanetal2014_enzo}
{Bryan} G.~L.,  et~al., 2014, \mn@doi [\apjs] {10.1088/0067-0049/211/2/19},
  \href {http://adsabs.harvard.edu/abs/2014ApJS..211...19B} {211, 19}

\bibitem[\protect\citeauthoryear{{Cen}}{{Cen}}{2002}]{cen2002}
{Cen} R.,  2002, \mn@doi [\apjs] {10.1086/339805}, \href
  {http://adsabs.harvard.edu/abs/2002ApJS..141..211C} {141, 211}

\bibitem[\protect\citeauthoryear{{Clarke}}{{Clarke}}{1996}]{clarke1996}
{Clarke} D.~A.,  1996, \mn@doi [\apj] {10.1086/176730}, \href
  {http://adsabs.harvard.edu/abs/1996ApJ...457..291C} {457, 291}

\bibitem[\protect\citeauthoryear{{Clarke}}{{Clarke}}{2010}]{clarke2010}
{Clarke} D.~A.,  2010, \mn@doi [\apjs] {10.1088/0067-0049/187/1/119}, \href
  {http://adsabs.harvard.edu/abs/2010ApJS..187..119C} {187, 119}

\bibitem[\protect\citeauthoryear{{Cunningham}, {Frank}, {Varni{\`e}re},
  {Mitran}  \& {Jones}}{{Cunningham}
  et~al.}{2009}]{cunninghametal2009_astrobear}
{Cunningham} A.~J.,  {Frank} A.,  {Varni{\`e}re} P.,  {Mitran} S.,   {Jones}
  T.~W.,  2009, \mn@doi [\apjs] {10.1088/0067-0049/182/2/519}, \href
  {http://adsabs.harvard.edu/abs/2009ApJS..182..519C} {182, 519}

\bibitem[\protect\citeauthoryear{{D'Angelo} \& {Bodenheimer}}{{D'Angelo} \&
  {Bodenheimer}}{2013}]{dangelobodenheimer2013}
{D'Angelo} G.,  {Bodenheimer} P.,  2013, \mn@doi [\apj]
  {10.1088/0004-637X/778/1/77}, \href
  {http://adsabs.harvard.edu/abs/2013ApJ...778...77D} {778, 77}

\bibitem[\protect\citeauthoryear{Dubey et~al.,}{Dubey
  et~al.}{2014}]{dubeyetal2014}
Dubey A.,  et~al., 2014, \mn@doi [Journal of Parallel and Distributed
  Computing] {http://dx.doi.org/10.1016/j.jpdc.2014.07.001}, 74, 3217

\bibitem[\protect\citeauthoryear{{Duffell} \& {MacFadyen}}{{Duffell} \&
  {MacFadyen}}{2011}]{duffellmacfayden2011}
{Duffell} P.~C.,  {MacFadyen} A.~I.,  2011, \mn@doi [\apjs]
  {10.1088/0067-0049/197/2/15}, \href
  {http://adsabs.harvard.edu/abs/2011ApJS..197...15D} {197, 15}

\bibitem[\protect\citeauthoryear{{Duffin} \& {Pudritz}}{{Duffin} \&
  {Pudritz}}{2008}]{duffinetal2008}
{Duffin} D.~F.,  {Pudritz} R.~E.,  2008, \mn@doi [\mnras]
  {10.1111/j.1365-2966.2008.14026.x}, \href
  {http://adsabs.harvard.edu/abs/2008MNRAS.391.1659D} {391, 1659}

\bibitem[\protect\citeauthoryear{{Dullemond} \& {Turolla}}{{Dullemond} \&
  {Turolla}}{2000}]{dullemondturolla2000_vtef}
{Dullemond} C.~P.,  {Turolla} R.,  2000, \aap, \href
  {http://adsabs.harvard.edu/abs/2000A%26A...360.1187D} {360, 1187}

\bibitem[\protect\citeauthoryear{{Evans} \& {Hawley}}{{Evans} \&
  {Hawley}}{1988}]{evanshawley1988}
{Evans} C.~R.,  {Hawley} J.~F.,  1988, \mn@doi [\apj] {10.1086/166684}, \href
  {http://adsabs.harvard.edu/abs/1988ApJ...332..659E} {332, 659}

\bibitem[\protect\citeauthoryear{{Fromang}, {Hennebelle}  \&
  {Teyssier}}{{Fromang} et~al.}{2006}]{fromangetal2006_ramses}
{Fromang} S.,  {Hennebelle} P.,   {Teyssier} R.,  2006, \mn@doi [\aap]
  {10.1051/0004-6361:20065371}, \href
  {http://adsabs.harvard.edu/abs/2006A%26A...457..371F} {457, 371}

\bibitem[\protect\citeauthoryear{{Fryxell} et~al.,}{{Fryxell}
  et~al.}{2000}]{fryxelletal2000_flash}
{Fryxell} B.,  et~al., 2000, \mn@doi [\apjs] {10.1086/317361}, \href
  {http://adsabs.harvard.edu/abs/2000ApJS..131..273F} {131, 273}

\bibitem[\protect\citeauthoryear{{Gudiksen}, {Carlsson}, {Hansteen}, {Hayek},
  {Leenaarts}  \& {Mart{\'{\i}}nez-Sykora}}{{Gudiksen} et~al.}{2011}]{bifrost}
{Gudiksen} B.~V.,  {Carlsson} M.,  {Hansteen} V.~H.,  {Hayek} W.,  {Leenaarts}
  J.,   {Mart{\'{\i}}nez-Sykora} J.,  2011, \mn@doi [\aap]
  {10.1051/0004-6361/201116520}, \href
  {http://adsabs.harvard.edu/abs/2011A%26A...531A.154G} {531, A154}

\bibitem[\protect\citeauthoryear{{Haugboelle}, {Grassi}, {Frostholm Mogensen},
  {Wielandt}, {Larsen}, {Vaytet}, {Connelly}  \& {Bizzarro}}{{Haugboelle}
  et~al.}{2017}]{haugboelle+17}
{Haugboelle} T.,  {Grassi} T.,  {Frostholm Mogensen} T.,  {Wielandt} D.,
  {Larsen} K.~K.,  {Vaytet} N.~M.,  {Connelly} J.,   {Bizzarro} M.,  2017, LPI
  Contributions, \href {http://adsabs.harvard.edu/abs/2017LPICo1975.2025H}
  {1975, 2025}

\bibitem[\protect\citeauthoryear{{Haugb{\o}lle}, {Frederiksen}  \&
  {Nordlund}}{{Haugb{\o}lle} et~al.}{2013}]{haugbolleetal2013_ppcode}
{Haugb{\o}lle} T.,  {Frederiksen} J.~T.,   {Nordlund} A.,  2013, \mn@doi
  [Physics of Plasmas] {10.1063/1.4811384}, \href
  {http://adsabs.harvard.edu/abs/2013PhPl...20f2904H} {20, 062904}

\bibitem[\protect\citeauthoryear{{Hayek}, {Asplund}, {Carlsson}, {Trampedach},
  {Collet}, {Gudiksen}, {Hansteen}  \& {Leenaarts}}{{Hayek}
  et~al.}{2010}]{bifrostRT}
{Hayek} W.,  {Asplund} M.,  {Carlsson} M.,  {Trampedach} R.,  {Collet} R.,
  {Gudiksen} B.~V.,  {Hansteen} V.~H.,   {Leenaarts} J.,  2010, \mn@doi [\aap]
  {10.1051/0004-6361/201014210}, \href
  {http://adsabs.harvard.edu/abs/2010A%26A...517A..49H} {517, A49}

\bibitem[\protect\citeauthoryear{{Hayes} \& {Norman}}{{Hayes} \&
  {Norman}}{2003}]{hayesnorman2003}
{Hayes} J.~C.,  {Norman} M.~L.,  2003, \mn@doi [\apjs] {10.1086/374658}, \href
  {http://adsabs.harvard.edu/abs/2003ApJS..147..197H} {147, 197}

\bibitem[\protect\citeauthoryear{{Heinemann}, {Dobler}, {Nordlund}  \&
  {Brandenburg}}{{Heinemann} et~al.}{2006}]{Heinemann2006}
{Heinemann} T.,  {Dobler} W.,  {Nordlund} {\AA}.,   {Brandenburg} A.,  2006,
  \mn@doi [\aap] {10.1051/0004-6361:20053120}, \href
  {http://adsabs.harvard.edu/abs/2006A%26A...448..731H} {448, 731}

\bibitem[\protect\citeauthoryear{{Hopkins}}{{Hopkins}}{2015}]{hopkins2015_gizmo}
{Hopkins} P.~F.,  2015, \mn@doi [\mnras] {10.1093/mnras/stv195}, \href
  {http://adsabs.harvard.edu/abs/2015MNRAS.450...53H} {450, 53}

\bibitem[\protect\citeauthoryear{{Hubber}, {Rosotti}  \& {Booth}}{{Hubber}
  et~al.}{2018}]{hubberetal2018_gandalf}
{Hubber} D.~A.,  {Rosotti} G.~P.,   {Booth} R.~A.,  2018, \mn@doi [\mnras]
  {10.1093/mnras/stx2405}, \href
  {http://adsabs.harvard.edu/abs/2018MNRAS.473.1603H} {473, 1603}

\bibitem[\protect\citeauthoryear{{Hubeny}}{{Hubeny}}{2003}]{Hubeny2003}
{Hubeny} I.,  2003, in {Hubeny} I.,  {Mihalas} D.,   {Werner} K.,  eds,
  Astronomical Society of the Pacific Conference Series Vol. 288, Stellar
  Atmosphere Modeling. p.~17

\bibitem[\protect\citeauthoryear{{Jiang}, {Stone}  \& {Davis}}{{Jiang}
  et~al.}{2012}]{Jiang2012}
{Jiang} Y.-F.,  {Stone} J.~M.,   {Davis} S.~W.,  2012, \mn@doi [\apjs]
  {10.1088/0067-0049/199/1/14}, \href
  {http://adsabs.harvard.edu/abs/2012ApJS..199...14J} {199, 14}

\bibitem[\protect\citeauthoryear{Kale, Bohm, Mendes, Wilmarth  \& Zheng}{Kale
  et~al.}{2008}]{kaleetal2008_charm}
Kale L.~V.,  Bohm E.,  Mendes C.~L.,  Wilmarth T.,   Zheng G.,  2008, in Bader
  D.,  ed., , Petascale Computing: Algorithms and Applications.
Chapman \& Hall / CRC Press, pp 421--441

\bibitem[\protect\citeauthoryear{{Klein}}{{Klein}}{1999}]{klein1999_orion}
{Klein} R.~I.,  1999, Journal of Computational and Applied Mathematics, \href
  {http://adsabs.harvard.edu/abs/1999JCoAM.109..123K} {109, 123}

\bibitem[\protect\citeauthoryear{{Kley}}{{Kley}}{1998}]{kley1998}
{Kley} W.,  1998, \aap, \href
  {http://adsabs.harvard.edu/abs/1998A%26A...338L..37K} {338, L37}

\bibitem[\protect\citeauthoryear{{Kravtsov}, {Klypin}  \&
  {Khokhlov}}{{Kravtsov} et~al.}{1997}]{kravtsovetal1997_art}
{Kravtsov} A.~V.,  {Klypin} A.~A.,   {Khokhlov} A.~M.,  1997, \mn@doi [\apjs]
  {10.1086/313015}, \href {http://adsabs.harvard.edu/abs/1997ApJS..111...73K}
  {111, 73}

\bibitem[\protect\citeauthoryear{{Kritsuk} et~al.,}{{Kritsuk}
  et~al.}{2011}]{kritsuketal2011}
{Kritsuk} A.~G.,  et~al., 2011, \mn@doi [\apj] {10.1088/0004-637X/737/1/13},
  \href {http://adsabs.harvard.edu/abs/2011ApJ...737...13K} {737, 13}

\bibitem[\protect\citeauthoryear{{Kuffmeier}, {Frostholm Mogensen},
  {Haugb{\o}lle}, {Bizzarro}  \& {Nordlund}}{{Kuffmeier}
  et~al.}{2016}]{kuffmeieretal2016_al26}
{Kuffmeier} M.,  {Frostholm Mogensen} T.,  {Haugb{\o}lle} T.,  {Bizzarro} M.,
  {Nordlund} {\AA}.,  2016, \mn@doi [\apj] {10.3847/0004-637X/826/1/22}, \href
  {http://adsabs.harvard.edu/abs/2016ApJ...826...22K} {826, 22}

\bibitem[\protect\citeauthoryear{{Kuffmeier}, {Haugb{\o}lle}  \&
  {Nordlund}}{{Kuffmeier} et~al.}{2017}]{kuffmeier+17}
{Kuffmeier} M.,  {Haugb{\o}lle} T.,   {Nordlund} {\AA}.,  2017, \mn@doi [\apj]
  {10.3847/1538-4357/aa7c64}, \href
  {http://adsabs.harvard.edu/abs/2017ApJ...846....7K} {846, 7}

\bibitem[\protect\citeauthoryear{{Levermore} \& {Pomraning}}{{Levermore} \&
  {Pomraning}}{1981}]{levermorepomraning1981_fld}
{Levermore} C.~D.,  {Pomraning} G.~C.,  1981, \mn@doi [\apj] {10.1086/159157},
  \href {http://adsabs.harvard.edu/abs/1981ApJ...248..321L} {248, 321}

\bibitem[\protect\citeauthoryear{{Lodato} \& {Price}}{{Lodato} \&
  {Price}}{2010}]{lodatoprice2010_phantom}
{Lodato} G.,  {Price} D.~J.,  2010, \mn@doi [\mnras]
  {10.1111/j.1365-2966.2010.16526.x}, \href
  {http://adsabs.harvard.edu/abs/2010MNRAS.405.1212L} {405, 1212}

\bibitem[\protect\citeauthoryear{{Mac Low}, {Norman}, {Konigl}  \&
  {Wardle}}{{Mac Low} et~al.}{1995}]{maclow+1995}
{Mac Low} M.-M.,  {Norman} M.~L.,  {Konigl} A.,   {Wardle} M.,  1995, \mn@doi
  [\apj] {10.1086/175477}, \href
  {http://adsabs.harvard.edu/abs/1995ApJ...442..726M} {442, 726}

\bibitem[\protect\citeauthoryear{{Masson}, {Teyssier}, {Mulet-Marquis},
  {Hennebelle}  \& {Chabrier}}{{Masson} et~al.}{2012}]{Massonetal2012}
{Masson} J.,  {Teyssier} R.,  {Mulet-Marquis} C.,  {Hennebelle} P.,
  {Chabrier} G.,  2012, \mn@doi [\apjs] {10.1088/0067-0049/201/2/24}, \href
  {http://adsabs.harvard.edu/abs/2012ApJS..201...24M} {201, 24}

\bibitem[\protect\citeauthoryear{{Mendygral} et~al.,}{{Mendygral}
  et~al.}{2017}]{wombat_2017}
{Mendygral} P.~J.,  et~al., 2017, \mn@doi [\apjs] {10.3847/1538-4365/aa5b9c},
  \href {http://adsabs.harvard.edu/abs/2017ApJS..228...23M} {228, 23}

\bibitem[\protect\citeauthoryear{{Meyer}, {Balsara}  \& {Aslam}}{{Meyer}
  et~al.}{2014}]{meyeretal2014_sts}
{Meyer} C.~D.,  {Balsara} D.~S.,   {Aslam} T.~D.,  2014, \mn@doi [Journal of
  Computational Physics] {10.1016/j.jcp.2013.08.021}, \href
  {http://adsabs.harvard.edu/abs/2014JCoPh.257..594M} {257, 594}

\bibitem[\protect\citeauthoryear{{Mignone}, {Bodo}, {Massaglia}, {Matsakos},
  {Tesileanu}, {Zanni}  \& {Ferrari}}{{Mignone}
  et~al.}{2007}]{mignoneetal2007_pluto}
{Mignone} A.,  {Bodo} G.,  {Massaglia} S.,  {Matsakos} T.,  {Tesileanu} O.,
  {Zanni} C.,   {Ferrari} A.,  2007, \mn@doi [\apjs] {10.1086/513316}, \href
  {http://adsabs.harvard.edu/abs/2007ApJS..170..228M} {170, 228}

\bibitem[\protect\citeauthoryear{{Nordlund}}{{Nordlund}}{1982}]{nordlund82}
{Nordlund} A.,  1982, \aap, \href
  {http://adsabs.harvard.edu/abs/1982A%26A...107....1N} {107, 1}

\bibitem[\protect\citeauthoryear{{Nordlund}}{{Nordlund}}{1984}]{nordlund1984}
{Nordlund} A.,  1984, {Iterative solution of radiative transfer problems with
  spherical symmetry using a single-ray approximation}.
Cambridge University Press, pp 211--233

\bibitem[\protect\citeauthoryear{{Nordlund}, {Galsgaard}  \&
  {Stein}}{{Nordlund} et~al.}{1994}]{nordlund+1994}
{Nordlund} {\AA}.,  {Galsgaard} K.,   {Stein} R.~F.,  1994, in {Rutten} R.~J.,
  {Schrijver} C.~J.,  eds,  NATO Advanced Science Institutes (ASI) Series C
  Vol. 433, NATO Advanced Science Institutes (ASI) Series C. p.~471

\bibitem[\protect\citeauthoryear{{Nordlund}, {Haugb{\o}lle}, {K{\"u}ffmeier},
  {Padoan}  \& {Vasileiades}}{{Nordlund} et~al.}{2014}]{nordlund+14}
{Nordlund} {\AA}.,  {Haugb{\o}lle} T.,  {K{\"u}ffmeier} M.,  {Padoan} P.,
  {Vasileiades} A.,  2014, in {Booth} M.,  {Matthews} B.~C.,   {Graham} J.~R.,
  eds,  IAU Symposium Vol. 299, Exploring the Formation and Evolution of
  Planetary Systems. pp 131--135, \mn@doi{10.1017/S1743921313008107}

\bibitem[\protect\citeauthoryear{{Padoan}, {Zweibel}  \& {Nordlund}}{{Padoan}
  et~al.}{2000}]{Padoanetal2000}
{Padoan} P.,  {Zweibel} E.,   {Nordlund} {\AA}.,  2000, \mn@doi [\apj]
  {10.1086/309299}, \href {http://adsabs.harvard.edu/abs/2000ApJ...540..332P}
  {540, 332}

\bibitem[\protect\citeauthoryear{{Padoan}, {Pan}, {Haugb{\o}lle}  \&
  {Nordlund}}{{Padoan} et~al.}{2016}]{padoan+2016}
{Padoan} P.,  {Pan} L.,  {Haugb{\o}lle} T.,   {Nordlund} {\AA}.,  2016, \mn@doi
  [\apj] {10.3847/0004-637X/822/1/11}, \href
  {http://adsabs.harvard.edu/abs/2016ApJ...822...11P} {822, 11}

\bibitem[\protect\citeauthoryear{{Padoan}, {Haugb{\o}lle}, {Nordlund}  \&
  {Frimann}}{{Padoan} et~al.}{2017}]{padoan+2017}
{Padoan} P.,  {Haugb{\o}lle} T.,  {Nordlund} {\AA}.,   {Frimann} S.,  2017,
  \mn@doi [\apj] {10.3847/1538-4357/aa6afa}, \href
  {http://adsabs.harvard.edu/abs/2017ApJ...840...48P} {840, 48}

\bibitem[\protect\citeauthoryear{{Paxton}, {Bildsten}, {Dotter}, {Herwig},
  {Lesaffre}  \& {Timmes}}{{Paxton} et~al.}{2010}]{paxton2010}
{Paxton} B.,  {Bildsten} L.,  {Dotter} A.,  {Herwig} F.,  {Lesaffre} P.,
  {Timmes} F.,  2010, {MESA: Modules for Experiments in Stellar Astrophysics},
  Astrophysics Source Code Library (\mn@eprint {ascl} {1010.083})

\bibitem[\protect\citeauthoryear{{Popovas}, {Nordlund}, {Ramsey}  \&
  {Ormel}}{{Popovas} et~al.}{submitted}]{popovasetal2018_arXiv}
{Popovas} A.,  {Nordlund} {\AA}.,  {Ramsey} J.~P.,   {Ormel} C.~W.,
  {submitted}, preprint, \href
  {http://adsabs.harvard.edu/abs/2018arXiv180107707P} {} (\mn@eprint {arXiv}
  {1801.07707})

\bibitem[\protect\citeauthoryear{{Porth}, {Xia}, {Hendrix}, {Moschou}  \&
  {Keppens}}{{Porth} et~al.}{2014}]{porthetal2014_amrvac}
{Porth} O.,  {Xia} C.,  {Hendrix} T.,  {Moschou} S.~P.,   {Keppens} R.,  2014,
  \mn@doi [\apjs] {10.1088/0067-0049/214/1/4}, \href
  {http://adsabs.harvard.edu/abs/2014ApJS..214....4P} {214, 4}

\bibitem[\protect\citeauthoryear{{Ramsey} \& {Dullemond}}{{Ramsey} \&
  {Dullemond}}{2015}]{ramseydullemond15}
{Ramsey} J.~P.,  {Dullemond} C.~P.,  2015, \mn@doi [\aap]
  {10.1051/0004-6361/201424954}, \href
  {http://adsabs.harvard.edu/abs/2015A%26A...574A..81R} {574, A81}

\bibitem[\protect\citeauthoryear{{Ramsey}, {Clarke}  \&
  {Men'shchikov}}{{Ramsey} et~al.}{2012}]{rcm12}
{Ramsey} J.~P.,  {Clarke} D.~A.,   {Men'shchikov} A.~B.,  2012, \mn@doi [\apjs]
  {10.1088/0067-0049/199/1/13}, \href
  {http://adsabs.harvard.edu/abs/2012ApJS..199...13R} {199, 13}

\bibitem[\protect\citeauthoryear{{Ramsey}, {Haugb{\o}lle}  \&
  {Nordlund}}{{Ramsey} et~al.}{submitted}]{ramsey+2018}
{Ramsey} J.~P.,  {Haugb{\o}lle} T.,   {Nordlund} {\AA}.,  submitted, in
  ASTRONUM2017 Proceedings. Journal of Physics Conference Series

\bibitem[\protect\citeauthoryear{{Rijkhorst}, {Plewa}, {Dubey}  \&
  {Mellema}}{{Rijkhorst} et~al.}{2006}]{rijhorstetal2006_hybrid}
{Rijkhorst} E.-J.,  {Plewa} T.,  {Dubey} A.,   {Mellema} G.,  2006, \mn@doi
  [\aap] {10.1051/0004-6361:20053401}, \href
  {http://adsabs.harvard.edu/abs/2006A%26A...452..907R} {452, 907}

\bibitem[\protect\citeauthoryear{{Robitaille}}{{Robitaille}}{2011}]{robitaille2011}
{Robitaille} T.~P.,  2011, \mn@doi [\aap] {10.1051/0004-6361/201117150}, \href
  {http://adsabs.harvard.edu/abs/2011A%26A...536A..79R} {536, A79}

\bibitem[\protect\citeauthoryear{{Shiokawa}, {Cheng}, {Noble}  \&
  {Krolik}}{{Shiokawa} et~al.}{2017}]{patchwork_2017}
{Shiokawa} H.,  {Cheng} R.~M.,  {Noble} S.~C.,   {Krolik} J.~H.,  2017,
  preprint, \href {http://adsabs.harvard.edu/abs/2017arXiv170105610S} {}
  (\mn@eprint {arXiv} {1701.05610})

\bibitem[\protect\citeauthoryear{{Springel}}{{Springel}}{2005}]{springel2005_gadget2}
{Springel} V.,  2005, \mn@doi [\mnras] {10.1111/j.1365-2966.2005.09655.x},
  \href {http://adsabs.harvard.edu/abs/2005MNRAS.364.1105S} {364, 1105}

\bibitem[\protect\citeauthoryear{{Springel}}{{Springel}}{2010}]{springel2010_arepo}
{Springel} V.,  2010, \mn@doi [\mnras] {10.1111/j.1365-2966.2009.15715.x},
  \href {http://adsabs.harvard.edu/abs/2010MNRAS.401..791S} {401, 791}

\bibitem[\protect\citeauthoryear{{Stein} \& {Nordlund}}{{Stein} \&
  {Nordlund}}{2012}]{stein+nordlundAR2012}
{Stein} R.~F.,  {Nordlund} {\AA}.,  2012, \mn@doi [\apjl]
  {10.1088/2041-8205/753/1/L13}, \href
  {http://adsabs.harvard.edu/abs/2012ApJ...753L..13S} {753, L13}

\bibitem[\protect\citeauthoryear{{Stone} \& {Norman}}{{Stone} \&
  {Norman}}{1992}]{stonenorman1992_zeus}
{Stone} J.~M.,  {Norman} M.~L.,  1992, \mn@doi [\apjs] {10.1086/191680}, \href
  {http://adsabs.harvard.edu/abs/1992ApJS...80..753S} {80, 753}

\bibitem[\protect\citeauthoryear{{Stone}, {Mihalas}  \& {Norman}}{{Stone}
  et~al.}{1992}]{stonetal1992}
{Stone} J.~M.,  {Mihalas} D.,   {Norman} M.~L.,  1992, \mn@doi [\apjs]
  {10.1086/191682}, \href {http://adsabs.harvard.edu/abs/1992ApJS...80..819S}
  {80, 819}

\bibitem[\protect\citeauthoryear{{Stone}, {Gardiner}, {Teuben}, {Hawley}  \&
  {Simon}}{{Stone} et~al.}{2008}]{stoneetal08}
{Stone} J.~M.,  {Gardiner} T.~A.,  {Teuben} P.,  {Hawley} J.~F.,   {Simon}
  J.~B.,  2008, \mn@doi [\apjs] {10.1086/588755}, \href
  {http://adsabs.harvard.edu/abs/2008ApJS..178..137S} {178, 137}

\bibitem[\protect\citeauthoryear{{Teyssier}}{{Teyssier}}{2002}]{teyssier2002_ramses}
{Teyssier} R.,  2002, \mn@doi [\aap] {10.1051/0004-6361:20011817}, \href
  {http://adsabs.harvard.edu/abs/2002A%26A...385..337T} {385, 337}

\bibitem[\protect\citeauthoryear{{Ziegler}}{{Ziegler}}{1998}]{ziegler1998_nirvana}
{Ziegler} U.,  1998, \mn@doi [Computer Physics Communications]
  {10.1016/S0010-4655(98)00022-8}, \href
  {http://adsabs.harvard.edu/abs/1998CoPhC.109..111Z} {109, 111}

\bibitem[\protect\citeauthoryear{{de Val-Borro} et~al.}{{de Val-Borro}
  et~al.}{2006}]{devalborroetal2006}
{de Val-Borro} M.,  et~al., 2006, \mn@doi [\mnras]
  {10.1111/j.1365-2966.2006.10488.x}, \href
  {http://adsabs.harvard.edu/abs/2006MNRAS.370..529D} {370, 529}

\bibitem[\protect\citeauthoryear{{van der Holst} et~al.,}{{van der Holst}
  et~al.}{2011}]{vanderholst2011_crash}
{van der Holst} B.,  et~al., 2011, \mn@doi [\apjs]
  {10.1088/0067-0049/194/2/23}, \href
  {http://adsabs.harvard.edu/abs/2011ApJS..194...23V} {194, 23}

\makeatother
\end{thebibliography}

%\begin{thebibliography}{99}
%\end{thebibliography}

%%%%%%%%%%%%%%%%%%%%%%%%%%%%%%%%%%%%%%%%%%%%%%%%%%

%%%%%%%%%%%%%%%%% APPENDICES %%%%%%%%%%%%%%%%%%%%%

% Don't change these lines
\bsp	% typesetting comment
\label{lastpage}
\end{document}